\documentclass[table]{article} %
\usepackage{iclr2025_conference,times}

\usepackage{amsmath,amsfonts,bm}

\def\eqref#1{equation~\ref{#1}}

\def\1{\bm{1}}

\DeclareMathAlphabet{\mathsfit}{\encodingdefault}{\sfdefault}{m}{sl}
\SetMathAlphabet{\mathsfit}{bold}{\encodingdefault}{\sfdefault}{bx}{n}

\usepackage[utf8]{inputenc} %
\usepackage[T1]{fontenc}    %
\usepackage{url}            %
\usepackage{booktabs}       %
\usepackage{amsfonts}       %
\usepackage{nicefrac}       %
\usepackage{microtype}      %
\usepackage{xspace}
\usepackage{graphicx}
\usepackage{makecell}
\usepackage{threeparttable}
\usepackage{adjustbox}
\usepackage{multicol}
\usepackage{multirow}
\usepackage{algorithm}
\usepackage{scalerel} %
\usepackage{enumitem}
\usepackage{wrapfig}
\usepackage{caption}

\usepackage[frozencache,cachedir=.]{minted}

\makeatletter
\DeclareRobustCommand\onedot{\futurelet\@let@token\@onedot}
\def\@onedot{\ifx\@let@token.\else.\null\fi\xspace}

\def\eg{\emph{e.g}\onedot} 
\def\ie{\emph{i.e}\onedot} 
 
\def\etc{\emph{etc}\onedot}

\makeatother

\definecolor{Gray}{gray}{0.97}
\definecolor{lightgray}{gray}{0.9}
\definecolor{lightblue}{rgb}{0.8,0.85,1}

\usepackage{bbding}

\newcommand{\no}{{\textcolor{red}{\XSolidBrush}}}
\definecolor{ao(english)}{rgb}{0.0, 0.5, 0.0}

\newcommand{\yes}{{\textcolor{ao(english)}{\CheckmarkBold}}}

\newcommand{\projname}{OpenHands\xspace}

\newcommand{\sref}[1]{\S\ref{#1}}
\newcommand{\fref}[1]{Fig.~\ref{#1}}
\newcommand{\tref}[1]{Tab.~\ref{#1}}

\NewDocumentCommand{\xingyao}
{ mO{} }{\textcolor{orange}{\textsuperscript{\textit{Xingyao}}\textsf{\textbf{\small[#1]}}}}

\NewDocumentCommand{\jas}
{ mO{} }{\textcolor{cyan}{\textsuperscript{\textit{Jas}}\textsf{\textbf{\small[#1]}}}}

\NewDocumentCommand{\hao}
{ mO{} }{\textcolor{blue}{\textsuperscript{\textit{Hao}}\textsf{\textbf{\small[#1]}}}}

\NewDocumentCommand{\gncomment}
{ mO{} }{\textcolor{red}{\textsuperscript{\textit{Graham}}\textsf{\textbf{\small[#1]}}}}

\NewDocumentCommand{\heng}
{ mO{} }{\textcolor{red}{\textsuperscript{\textit{Heng}}\textsf{\textbf{\small[#1]}}}}

\NewDocumentCommand{\hoang}
{ mO{} }{\textcolor{green}{\textsuperscript{\textit{Hoang}}\textsf{\textbf{\small[#1]}}}}

\NewDocumentCommand{\fx}
{ mO{} }{\textcolor{yellow}{\textsuperscript{\textit{Frank}}\textsf{\textbf{\small[#1]}}}}

\NewDocumentCommand{\jiayi}
{ mO{} }{\textcolor{brown}{\textsuperscript{\textit{Jiayi}}\textsf{\textbf{\small[#1]}}}}

\NewDocumentCommand{\binyuan}
{ mO{} }{\textcolor{blue}{\textsuperscript{\textit{Binyuan}}\textsf{\textbf{\small[#1]}}}}

\newcommand{\slack}{\raisebox{-1.5pt}{\includegraphics[height=1.05em]{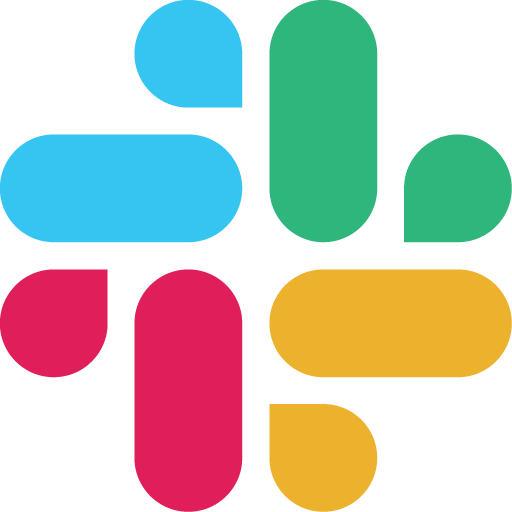}}\xspace}
\newcommand{\github}{\raisebox{-1.5pt}{\includegraphics[height=1.05em]{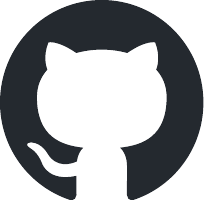}}\xspace}

\newcommand{\numbenchmarks}{15\xspace}

\usepackage{url}
\definecolor{scholarblue}{rgb}{0.21,0.49,0.74}
\usepackage[pagebackref,breaklinks,colorlinks]{hyperref}
\hypersetup{
	colorlinks=true,
	linkcolor=red,
	filecolor=blue,      
	urlcolor=scholarblue,
	citecolor=scholarblue,
}

\title{\raisebox{-3pt}{\includegraphics[height=1.2em]{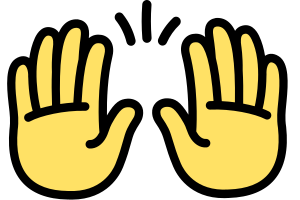}}\xspace
\projname: An Open Platform for \\
AI Software Developers as Generalist Agents}

\author{Xingyao Wang$^{1,10}$, Boxuan Li$^2$, Yufan Song$^2$, Frank F. Xu$^2$, Xiangru Tang$^3$, \\
\textbf{Mingchen Zhuge$^6$, Jiayi Pan$^4$, Yueqi Song$^2$, Bowen Li, Jaskirat Singh$^7$,} \\
\textbf{Hoang H. Tran$^8$, Fuqiang Li, Ren Ma, Mingzhang Zheng, Bill Qian$^3$, Yanjun Shao$^3$,}\\
\textbf{Niklas Muennighoff\hspace{1pt}$^5$, Yizhe Zhang, Binyuan Hui$^{9}$, Junyang Lin$^{9}$,} \\
\textbf{Robert Brennan$^{10}$, Hao Peng$^1$, Heng Ji$^1$, Graham Neubig$^{2,10}$}
\\
$^1$UIUC \ \
$^2$CMU \ \
$^3$Yale \ \
$^4$UC Berkeley \ 
$^5$Contextual AI \ \
$^6$KAUST \ \
$^7$ANU \ \\
$^8$HCMUT \ \
$^{9}$Alibaba \ \
$^{10}$All Hands AI\\
\texttt{xingyao6@illinois.edu, gneubig@cs.cmu.edu}
}

\iclrfinalcopy %
\begin{document}

\maketitle

\begin{abstract}
Software is one of the most powerful tools that we humans have at our disposal; it allows a skilled programmer to interact with the world in complex and profound ways.
At the same time, thanks to improvements in large language models (LLMs), there has also been a rapid development in AI agents that interact with and affect change in their surrounding environments.
In this paper, we introduce \projname (\emph{f.k.a.} OpenDevin), a platform for the development of powerful and flexible AI agents that interact with the world in similar ways to those of a human developer: by writing code, interacting with a command line, and browsing the web.
We describe how the platform allows for the implementation of new agents, safe interaction with sandboxed environments for code execution, coordination between multiple agents, and incorporation of evaluation benchmarks.
Based on our currently incorporated benchmarks, we perform an evaluation of agents over \numbenchmarks challenging tasks, including software engineering (\eg, \textsc{Swe-bench}) and web browsing (\eg, \textsc{WebArena}), among others.
Released under the permissive MIT license, \projname is a community project spanning academia and industry with more than 2.1K contributions from over 188 contributors.

\begin{center}
\begin{tabular}{rll}
    \github & \textbf{\small{Code}} & \url{https://github.com/All-Hands-AI/OpenHands}\\
    \slack & \textbf{\small{Slack}} & \url{http://bit.ly/OpenHands-Slack}\\
\end{tabular}
\end{center}

\end{abstract}

\section{Introduction}

Powered by large language models (LLMs; \citealt{openai2024gpt4o,team2023gemini,jiang2024mixtral,chang2024survey}), user-facing AI systems (such as ChatGPT) have become increasingly capable of performing complex tasks such as accurately responding to user queries, solving math problems, and generating code.
In particular, AI \emph{agents}, systems that can perceive and act upon the external environment, have recently received ever-increasing research focus.
They are moving towards performing complex tasks such as developing software \citep{jimenez2024swebench},
navigating real-world websites \citep{zhou2023webarena}, doing household chores \citep{ahn2022can}, or even performing scientific research \citep{boiko2023autonomous,tang2024prioritizing}. 

As AI agents become capable of tackling complex problems, their development and evaluation have also become challenging. 
There are numerous recent efforts in creating open-source frameworks that facilitate the development of agents \citep{hong2023metagpt,chen2024autoagents,wu2023autogen}.
These agent frameworks generally include: 1) \textbf{interfaces} through which agents interact with the world (such as JSON-based function calls or code execution), 2) \textbf{environments} in which agents operate, and 3) \textbf{interaction mechanisms} for human-agent or agent-agent communication.
These frameworks streamline and ease the development process in various ways (\tref{tab:comparison}, \sref{sec:related_work}).

When designing AI agents, we can also consider how \emph{human} interacts with the world.
The most powerful way in which humans currently interact with the world is through \emph{software} -- software powers every aspect of our life, supporting everything from the logistics for basic needs to the advancement of science, technology, and AI itself.
Given the power of software, as well as the existing tooling around its efficient development, use, and deployment, it provides the ideal interface for AI agents to interact with the world in complex ways.
However, building agents that can effectively develop software comes with its own unique challenges.
How can we enable agents to effectively \emph{create and modify code in complex software systems}?
How can we provide them with tools to \emph{gather information on-the-fly} to debug problems or gather task-requisite information?
How can we ensure that development is \emph{safe and avoids negative side effects} on the users' systems?

In this paper, we introduce \projname (\emph{f.k.a.} OpenDevin), a community-driven platform designed for the development of generalist and specialist AI agents that interact with the world through software.%
\footnote{While initially inspired by AI software engineer Devin \citep{cognition_devin}, \projname has quickly evolved to support much wider range of applications beyond software engineering through diverse community contributions.}
It features:
\begin{itemize}[noitemsep,topsep=0pt,parsep=2pt,partopsep=0pt,leftmargin=18pt]
\item[(1)] An \textbf{interaction mechanism} which allows user interfaces, agents, and environments to interact through an \emph{event stream} architecture that is powerful and flexible (\sref{sec:agent-abstraction}).
\item[(2)] A \textbf{runtime environment} that consists of a docker-sandboxed operating system with a bash shell, a web browser, and IPython server that the agents can interact with (\sref{sec:agent-runtime}).
\item[(3)] An \textbf{interface} allowing the agent to interact with the environment in a manner similar to actual software engineers (\sref{sec:agent-skills}). We provide the capability for agents to a) create and edit complex software, b) execute arbitrary code in the sandbox, and c) browse websites to collect information.
\item[(4)] \textbf{Multi-agent delegation}, allowing multiple specialized agents to work together (\sref{sec:agent-delegation}).
\item[(5)] \textbf{Evaluation framework}, facilitating the evaluation of agents across a wide range of tasks (\sref{sec:evaluation}).

\end{itemize}

\begin{figure}
\centering
\vspace{-0.2cm}
\includegraphics[clip, trim=4.5cm 4.5cm 4.5cm 4.6cm, width=\textwidth]{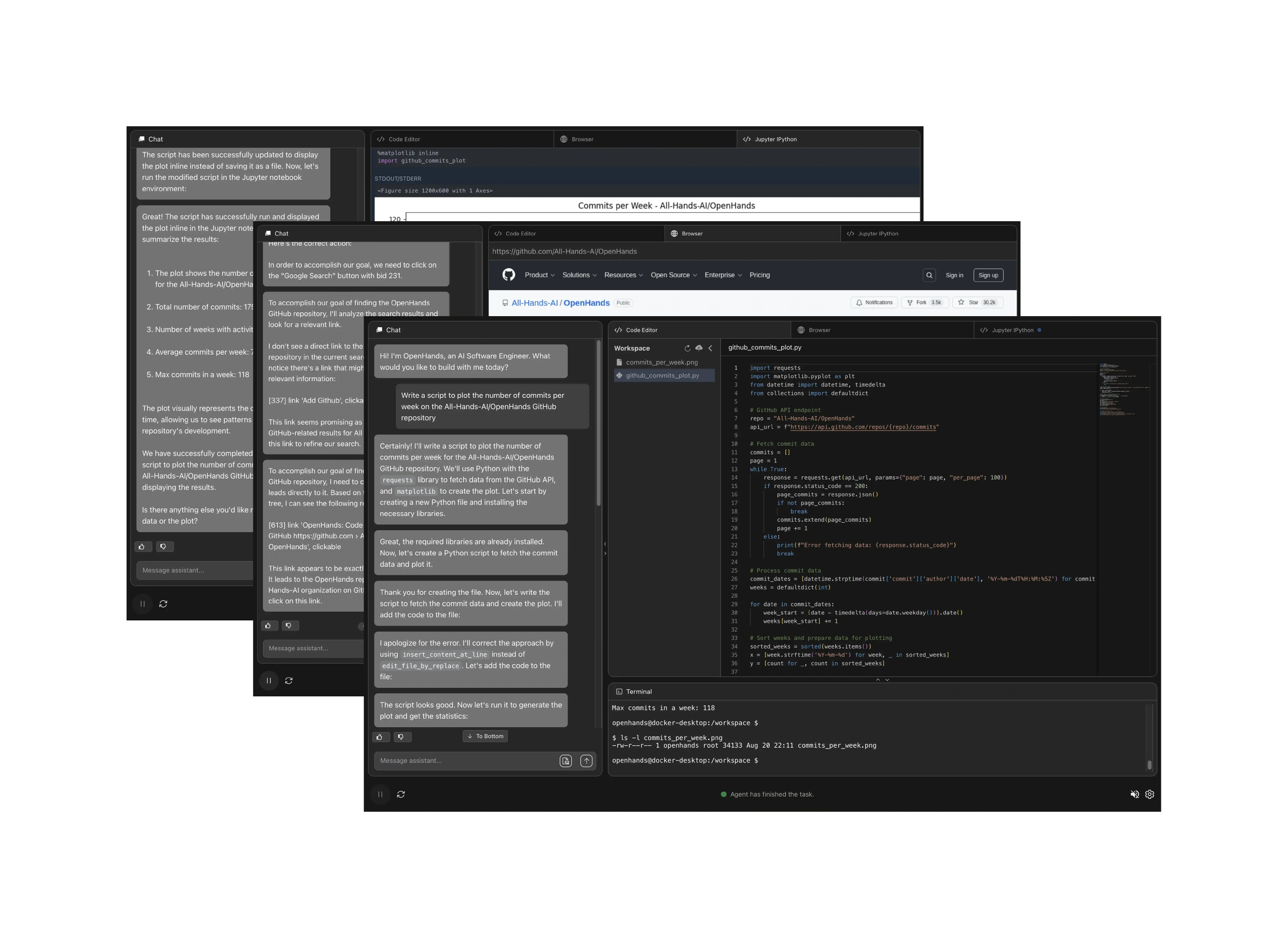}
\vspace{-0.6cm}
\caption{
\projname User Interface (UI, \sref{sec:ui}) allows users to view files, check executed bash commands/Python code, observe the agent's browser activity, and directly interact with the agent.
}
\label{fig:ui}
\vskip -0.1in
\end{figure}

Importantly, \projname is not just a conceptual framework, but it also includes a comprehensive and immediately usable implementation of agents, environments, and evaluations.
As of this writing, \projname includes an agent hub with over 10 implemented agents (\sref{sec:agenthub}), including a strong generalist agent implemented based on the CodeAct architecture \citep{wang2024executable}, with additions for web browsing \citep{browsergym} and code editing specialists \citep{yang2024sweagent}.
Interaction with users is implemented through a chat-based user interface that visualizes the agent's current actions and allows for real-time feedback (\fref{fig:ui}, \sref{sec:ui}).
Furthermore, the evaluation framework currently supports \numbenchmarks benchmarks, which we use to evaluate our agents (\sref{sec:evaluation}).

Released under a permissive MIT license allowing commercial use,
\projname is poised to support a diverse array of research and real-world applications across academia and industry.
\projname has gained significant traction, with 32K GitHub stars and more than 2.1K contributions from over 188 contributors.
We envision \projname as a catalyst for future research innovations and diverse applications driven by a broad community of practitioners.

\section{\projname Architecture}

\begin{figure}[t]
\vspace{-10pt}
\centering
\includegraphics[clip, trim=4.5cm 4.5cm 4.5cm 4.5cm, width=\textwidth]{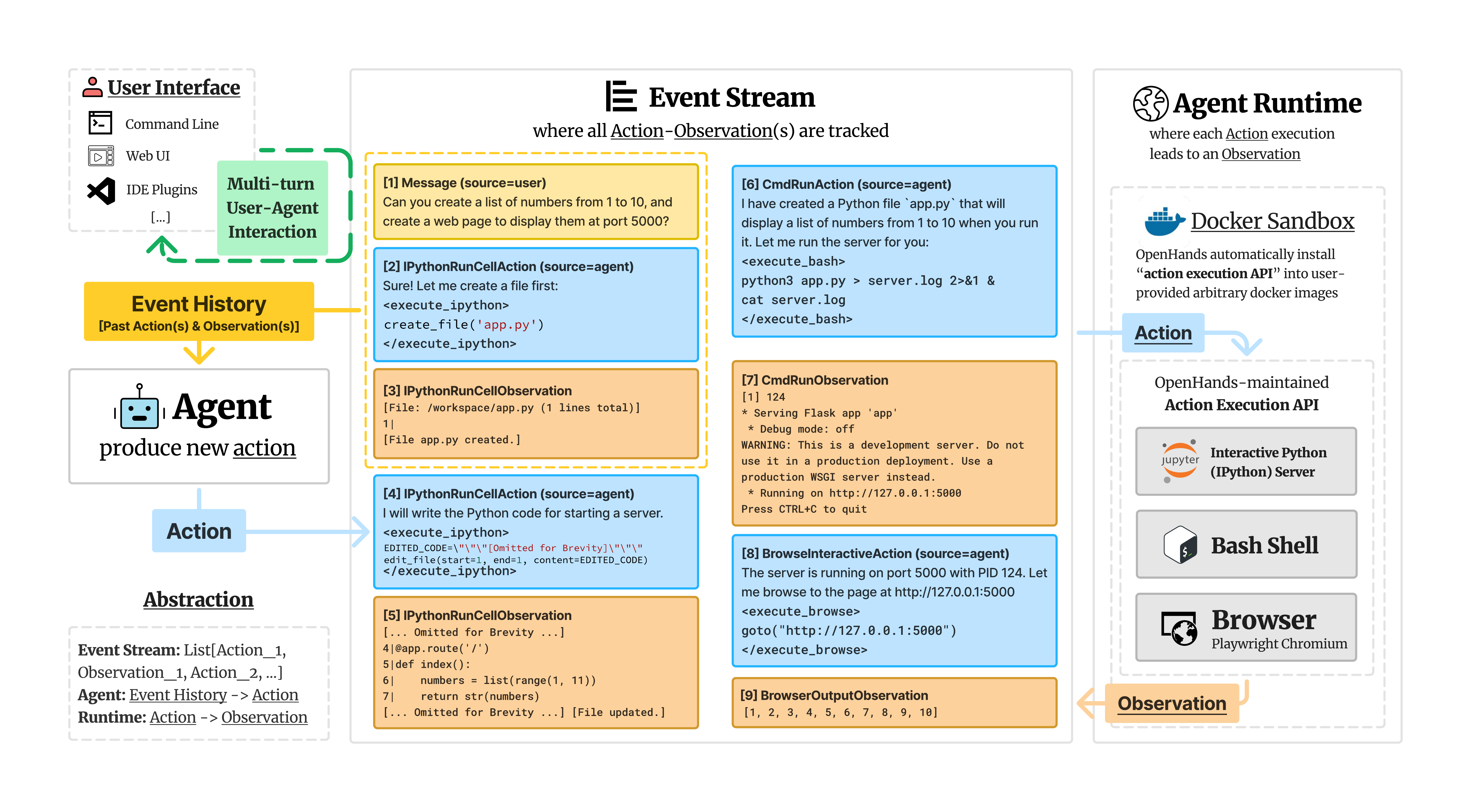}
\caption{
\projname consists of 3 main components:
1) \textbf{Agent abstraction} where community can contribute different implementation of agents (\sref{sec:agent-abstraction}) into agenthub (\sref{sec:agenthub});
2) \textbf{Event stream} for tracking history of actions and observations; 
3) \textbf{Runtime} to execute all actions into observations (\sref{sec:agent-runtime}).
}
\vskip -0.15in
\label{fig:arch}
\end{figure}

We next describe using \projname in detail. In particular, we discuss 1) how to define and implement an agent (\sref{sec:agent-abstraction}), 2) how each action execution leads to an observation (\sref{sec:agent-runtime}), 3) how to reliably manage and extend commonly used skills for agents (\sref{sec:agent-skills}), and 4) how to compose multiple agents together for task solving (\sref{sec:agent-delegation}).
\fref{fig:arch} provides an overview.

\subsection{Agent Definition and Implementation}
\label{sec:agent-abstraction}

An \textbf{agent} can perceive the \textbf{state} of the environment (\eg, prior actions and observations) and produce an \textbf{action} for execution while solving a user-specified task.

\noindent \textbf{The State and Event Stream.} In \projname, the state is a data structure that encapsulates all relevant information for the agent's execution. A key component of this state is the \textbf{event stream}, which is a chronological collection of past actions and observations, including the agent's own actions and user interactions (\eg, instructions, feedback). In addition to the event stream, the state incorporates auxiliary information for agent's operation, such as the accumulative cost of LLM calls, metadata to track multi-agent delegation (\sref{sec:agent-delegation}), and other execution-related parameters.

\noindent \textbf{Actions.} Inspired by CodeAct \citep{wang2024executable}, \projname connects an agent with the environment through a core set of general actions.
Actions \texttt{IPythonRunCellAction} and \texttt{CmdRunAction} enable the agent to execute \textit{arbitrary} Python code and bash commands inside the sandbox environment (\eg, a securely isolated Linux operating system).
\texttt{BrowserInteractiveAction} enables interaction with a web browser with a domain-specific language for browsing introduced by BrowserGym \citep{workarena2024}.
These actions were chosen to provide a comprehensive yet flexible set of primitives covering most tasks performed by human software engineers and analysts.
The action space based on programming languages (PL) is powerful and flexible enough to perform any task with tools in different forms (\eg, Python function, REST API, \etc) while being reliable and easy to maintain \citep{wang2024executable} .

\begin{wrapfigure}[21]{R}{0.6\textwidth}
\begin{minipage}{0.6\textwidth}
\vspace{-0.4cm}
\caption{Minimal example of implementing an agent in \projname.}
\vspace{-0.4cm}
\label{algo:agent-abstraction}
\begin{minted}[frame=lines,fontsize=\tiny]{python}
class MinimalAgent:
    def reset(self) -> None:
        self.system_message = "You are a helpful assistant ..."

    def step(self, state: State):
        messages: list[dict[str, str]] = [
            {'role': 'system', 'content': self.system_message}
        ]
        for prev_action, obs in state.history:
            action_message = get_action_message(prev_action)
            messages.append(action_message)
            obs_message = get_observation_message(obs)
            messages.append(obs_message)

        # use llm to generate response (e.g., thought, action)
        response = self.llm.do_completion(messages)

        # parse and execute action in the runtime
        action = self.parse_response(response)
        if self.is_finish_command(action):
            return AgentFinishAction()
        elif self.is_bash_command(action):
            return CmdRunAction(command=action.command)
        elif self.is_python_code(action):
            return IPythonRunCellAction(code=action.code)
        elif self.is_browser_action(action):
            return BrowseInteractiveAction(code=action.code)
        else:
            return MessageAction(content=action.message)
\end{minted}
\end{minipage}
\vspace{-2cm}
\end{wrapfigure}

This design is also compatible with existing tool-calling agents that require a list of pre-defined tools \citep{langchain2022}. That is, users can easily define tools using PL supported in primitive actions (\eg, write a Python function for calculator) and make those tools available to the agent through JSON-style function-calling experiences \citep{toolbench}.
Moreover, the framework's powerful PL-based primitives further make it possible for the agents to create tools by themselves (\eg, by generating Python functions, \citealt{craft}) when API to complete the task is unavailable.
Refer to \sref{sec:agent-skills} for how these core PL-based actions can be composed into a diverse set of tools.

\noindent \textbf{Observations.} Observations describe the environmental changes (\eg, execution result of prior actions, text messages from the human user \etc) that the agent observes. 

\noindent \textbf{Implement a New Agent.} 
The agent abstraction is designed to be simple yet powerful, allowing users to create and customize agents for various tasks easily.
The core of the agent abstraction lies in the \texttt{step} function, which takes the current state as input and generates an appropriate action based on the agent's logic. Simplified example code for the agent abstraction is illustrated in \fref{algo:agent-abstraction}.
By providing this abstraction, \projname allows the users to focus on defining desired agent behavior and logic without worrying about the low-level details of how actions are executed (\sref{sec:agent-runtime}).

\subsection{Agent Runtime: How Execution of Actions Results in Observations}
\label{sec:agent-runtime}

Agent Runtime provides a general environment that equips the agent with an action space comparable to that of human software developers, enabling \projname agents to tackle a wide range of software development and web-based tasks, including complex software development workflows, data analysis projects, web browsing tasks, and more.
It allows the agent to access a bash terminal to run code and command line tools, utilize a Jupyter notebook for writing and executing code on-the-fly, and interact with a web browser for web-based tasks (\eg, information seeking). 

\noindent \textbf{Docker Sandbox.} For each task session, \projname spins up a securely isolated docker container sandbox, where all the actions from the event stream are executed.
\projname connects to the sandbox through a REST API server running inside it (i.e., the \projname action execution API), executes arbitrary actions (e.g., bash command, python code) from the event stream, and returns the execution results as observations.
A configurable workspace directory containing files the user wants the agent to work on is mounted into that secure sandbox for \projname agents to access.

\noindent \textbf{\projname Action Execution API.} \projname maintains an API server that runs \textit{inside the docker sandbox} to listen for action execution requests from the event stream. The API server maintains:

\begin{itemize}[noitemsep,topsep=0pt,parsep=2pt,partopsep=0pt,leftmargin=18pt]
\item[(1)] A bash shell that connects with the operating system environment (specified by the docker image) for command execution.

\item[(2)] A Jupyter IPython server to handle interactive \emph{python} \citep{ipython} code execution requests and return the execution results back to the event stream.
\item[(3)] A Chromium browser based on \cite{playwright}. The provider provides a set of action primitives defined by BrowserGym \citep{browsergym, workarena2024}, such as navigation, clicking, typing, and scrolling. The full set of actions is detailed in \sref{sec:browsergym-actions}. 
After executing these actions, the browser runtime provides a rich set of observations about the current state of the browser, including HTML, DOM, accessibility tree~\citep{accessibilitytree}, screenshot, opened tabs, \etc.
\end{itemize}

\noindent \textbf{Arbitrary Docker Image Support.} \projname allows agents to run on arbitrary operating systems with different software environments by supporting runtime based on arbitrary docker images.
\projname implements a build mechanism that takes a user-provided arbitrary docker image and installs \projname action execution API into that image to allow for agent interactions.
We include a detailed description of \projname agent runtime in \sref{sec:runtime-details}.

\subsection{Agent Skills: The Extensible Agent-Computer Interface}
\label{sec:agent-skills}

SWE-Agent \citep{yang2024sweagent} highlights the importance of a carefully crafted Agent-Computer Interface (ACI, \ie, specialized tools for particular tasks) in successfully solving complex tasks.
However, creating, maintaining, and distributing a wide array of tools can be a daunting engineering challenge, especially when we want to make these tools available to different agent implementations (\sref{sec:agenthub}).
To tackle these, we build an \textbf{AgentSkills library}, a toolbox designed to enhance the capabilities of agents, offering utilities not readily available through basic \textit{bash} commands or \textit{python} code.

\textbf{Easy to create and extend tools.} AgentSkills is designed as a Python package consisting of different utility functions (\ie, tools) that are automatically imported into the Jupyter IPython environment (\sref{sec:agent-runtime}).
The ease of defining a Python function as a tool lowers the barrier for community members to contribute new tools to the library.
The generality of Python packages also allows different agent implementations to easily leverage these tools through one of our core action \texttt{IPythonRunCellAction} (\sref{sec:agent-abstraction}).

\textbf{Inclusion criteria and philosophy.} 
In the AgentSkills library, we do not aim to wrap every possible Python package and re-teach agents their usage (\eg, LLM already knows \texttt{pandas} library that can read CSV file, so we don't need to re-create a tool that teaches the agent to read the same file format).
We only add a new skill when: (1) it is not readily achievable for LLM to write code directly (\eg, edit code and replace certain lines), and/or (2) it involves calling an external model (\eg, calling a speech-to-text model, or model for code editing \citep{speculative_editing}).

\textbf{Currently supported skills.}
AgentSkills library includes file editing utilities adapted from SWE-Agent \citep{yang2024sweagent} and Aider \citep{aider} like \texttt{edit\_file}, which allows modifying an existing file from a specified line; scrolling functions \texttt{scroll\_up} and \texttt{scroll\_down} for viewing a different part of files.
It also contains tools that support reading multi-modal documents, like \texttt{parse\_image} and \texttt{parse\_pdf} for extracting information from images using vision-language models (\eg, GPT-4V) and reading text from PDFs, respectively.
A complete list of supported skills can be found in \sref{sec:supported-agentskills}.

\subsection{Agent Delegation: Cooperative Multi-agent Interaction}
\label{sec:agent-delegation}

\projname allows interactions between multiple agents as well. To this end, we use a special action type \texttt{AgentDelegateAction}, which enables an agent to delegate a specific subtask to another agent.
For example, the generalist CodeActAgent, with limited support for web-browsing, can use \texttt{AgentDelegateAction} to delegate web browsing tasks to the specialized BrowsingAgent to perform more complex browsing activity (\eg, navigate the web, click buttons, submit forms, \etc).

\begin{table}[!tbp]
\vskip -0.15in
\caption{
Comparison of different AI agent frameworks (\sref{sec:related_work}).
\textsc{Swe} refers to `software engineering'.
\textbf{Standardized tool library}: if framework contains reusable tools for different agent implementations (\sref{sec:agent-skills}); 
\textbf{Built-in sandbox \& code execution}: if it supports sandboxed execution of arbitrary agent-generated code;
\textbf{Built-in web browser}: if it provides agents access to a fully functioning web browser;
\textbf{Human-AI collaboration}: if it enables multi-turn human-AI collaboration (\eg, human can interrupt the agent during task execution and/or provide additional feedback and instructions);
\textbf{AgentHub}: if it hosts implementations of various agents (\sref{sec:agenthub});
\textbf{Evaluation Framework}: if it offers systematic evaluation of implemented agents on challenging benchmarks (\sref{sec:evaluation});
\textbf{Agent QC} (Quality Control): if the framework integrates tests (\sref{sec:integration-tests}) to ensure overall framework software quality.
}
\vspace{-0.2cm}
\label{tab:comparison}
\centering
\scriptsize
\begin{threeparttable}
\begin{adjustbox}{width=1\textwidth}
\begin{tabular}{ll|ccccccccc}
\toprule
\textbf{Framework} 
& \textbf{Domain} 
& \textbf{\makecell{Graphic\\ User Interface}}
& \textbf{\makecell{Standardized\\ Tool Library}}
& \textbf{\makecell{Built-in Sandbox\\ \& Code Execution}}
& \textbf{\makecell{Built-in Web\\ Browser}}
& \textbf{\makecell{Multi-agent\\ Collaboration}}
& \textbf{\makecell{Human-AI\\ Collaboration}} %
& \textbf{\makecell{AgentHub}} %
& \textbf{\makecell{Evaluation\\ Framework}}
& \textbf{\makecell{Agent\\ QC}} \\
\midrule

AutoGPT~\cite{gravitasauto} & General & \yes & \no & \no & \no & \no & \no & \yes & \no & \yes \\

LangChain~\citep{langchain2022} & General & \no & \yes & \no$^*$ & \no$^*$ & \no & \no & \yes & \no & \no \\
MetaGPT~\citep{hong2023metagpt} & General & \no & \yes & \no & \yes & \yes & \no & \yes & \no & \yes \\
AutoGen~\citep{wu2023autogen} & General & \no & \yes & \yes & \yes & \yes & \yes & \yes & \yes & \no \\
AutoAgents~\citep{chen2024autoagents} & General & \no & \no & \no & \no & \yes & \no & \no & \no & \no \\
Agents~\citep{zhou2023agents} & General & \no & \no & \no & \no & \yes & \yes & \no & \no & \no \\
Xagents~\citep{xagent2023} & General & \yes & \yes & \no & \yes & \yes & \no & \yes & \no & \no \\
OpenAgents~\citep{xie2023openagents} & General & \yes & \no & \yes & \yes & \no & \no & \yes & \no & \no \\
GPTSwarm~\citep{zhuge2024language} & General & \no & \yes & \no & \no & \yes & \yes & \no & \no & \no \\
\midrule
AutoCodeRover~\citep{zhang2024autocoderover} & SWE & \no & \no & \yes & \no & \no & \no & \no & \no & \no \\
SWE-Agent~\citep{yang2024sweagent} & SWE & \no & \no & \yes & \no & \no & \no & \no & \no & \no \\
\midrule
\textbf{\projname} & General & \yes & \yes & \yes & \yes & \yes & \yes & \yes & \yes & \yes \\
\bottomrule     
\end{tabular}
\end{adjustbox}
{
\tiny
\begin{tablenotes}
\item[*] No native support. Third-party commercial options are available.
\end{tablenotes}
}
\vskip -0.15in
\end{threeparttable}
\end{table}

\section{AgentHub: A Hub of Community-Contributed Agents}
\label{sec:agenthub}

Based on our agent abstraction (\sref{sec:agent-abstraction}), \projname supports a wide range of community-contributed agent implementations for end users to choose from and act as baselines for different agent tasks.

\noindent \textbf{CodeAct Agent.} CodeActAgent is the default generalist agent based on the CodeAct framework \citep{wang2024executable}. At each step, the agent can (1) converse to communicate with humans in natural language to ask for clarification, confirmation, \etc, or (2) to perform the task by executing code (\textit{a.k.a.}, \textbf{CodeAct}), including executing bash commands, Python code, or browser-specific programming language (\sref{sec:agent-runtime}). This general action space allows the agent (v1.5 and above) to perform various tasks, including editing files, browsing the web, running programs, etc.

\noindent \textbf{Browsing Agent.} 
We implemented a generalist web agent called Browsing Agent, to serve as a simple yet effective baseline for web agent tasks.
The agent is similar to that in WebArena~\citep{zhou2023webarena}, but with improved observations and actions, with only zero-shot prompting.
Full prompts are in~\sref{sec:browsing-agent-details}. 

\noindent \textbf{GPTSwarm Agent.} 
GPTSwarm~\citep{zhuge2024language} pioneers the use of optimizable graphs to construct agent systems, unifying language agent frameworks through modularity. Each node represents a distinct operation, while edges define collaboration and communication pathways. This design allows automatic optimization of nodes and edges, driving advancements in creating multi-agent systems.

\noindent \textbf{Micro Agent(s).}
In addition, \projname enables the creation of \textbf{micro agent}, an agent \textit{specialized} towards a particular task.
A micro agent re-uses most implementations from an existing generalist agent (e.g., CodeAct Agent). 
It is designed to lower the barrier to agent development, where community members can share specialized prompts that work well for their particular use cases.

\everypar{\looseness=-1}

\section{Evaluation}
\label{sec:evaluation}

\begin{wraptable}[13]{r}{0.58\textwidth}
\vspace{-0.4cm}
\caption{Evaluation benchmarks in \projname.}
\vspace{-0.2cm}
\begin{adjustbox}{max width=0.58\textwidth}
\begin{tabular}{lll}
\toprule
\textbf{Category}
& \textbf{Benchmark}
& \textbf{Required Capability} \\
\midrule
\multirow{7}{*}{\textbf{Software}}
& SWE-Bench \citep{jimenez2024swebench} & Fixing Github issues \\
& HumanEvalFix \citep{muennighoff2024octopack} & Fixing Bugs \\
& BIRD \citep{bird} & Text-to-SQL \\
& BioCoder \citep{tang2023biocoder} & Bioinformatics coding \\
& ML-Bench \citep{liu2024mlbench} & Machine learning coding \\
& Gorilla APIBench \citep{patil2023gorilla} & Software API calling \\
& ToolQA \citep{zhuang2024toolqa} & Tool use \\

\midrule
\multirow{2}{*}{\textbf{Web}}
& WebArena \citep{zhou2023webarena} & Goal planning \& realistic browsing \\
& MiniWoB++ \citep{liu2018reinforcement} & Short trajectory on synthetic web \\

\midrule
\multirow{6}{*}{\textbf{Misc.~Assistance}}
& GAIA \citep{gaia} & Tool-use, browsing, multi-modality \\
& GPQA \citep{rein2023gpqa} & Graduate-level Google-proof Q\&A \\
& AgentBench \citep{liu2023agentbench} & Operating system interaction (bash) \\
& MINT \citep{wang2024mint} & Multi-turn math and code problems \\
& Entity Deduction Arena \citep{zhang2023entity} & State tracking \& strategic planning \\
& ProofWriter \citep{proofwriter} & Deductive Logic Reasoning \\

\bottomrule
\vspace{-20mm}
\end{tabular}
\end{adjustbox}
\label{tab:od-benchmarks}
\end{wraptable}

To systematically track progress in building generalist digital agents, as listed in \tref{tab:od-benchmarks}, we integrate \numbenchmarks established benchmarks into \projname. These benchmarks cover software engineering, web browsing, and miscellaneous assistance.
In this section, we compare \projname to open-source reproducible baselines that do not perform manual prompt engineering specifically based on the benchmark \textit{content}.
Please note that we use `OH' as shorthand for \projname for the rest of this section for brevity reasons.

\subsection{Result Overview} 
In \projname, our goal is to develop \textbf{general digital agents} capable of interacting with the world through software interfaces (as exemplified by the code actions described in \sref{sec:agent-abstraction}). We recognize that a software agent should excel not only in code editing but also in web browsing and various auxiliary tasks, such as answering questions about code repositories or conducting online research.

\tref{tab:overall_perf} showcases a curated set of evaluation results. While \projname agents may not achieve top performance in every category, they are designed with generality in mind. Notably, the same CodeAct agent, without any modifications to its system prompt, demonstrates competitive performance across three major task categories: software development, web interaction, and miscellaneous tasks. This is particularly significant when compared to the baseline agents, which are typically designed and optimized for specific task categories.

{
\begin{table*}[!t]
\centering
\vspace{-0.5cm}
\caption{Selected evaluation results for \projname agents (\sref{sec:evaluation}).
See \tref{tab:swe_bench} (software), \tref{tab:web_browsing} (web), \tref{tab:misc_assistance} (miscellaneous assistance) for full results across benchmarks.}
\vspace{-0.2cm}
\begin{adjustbox}{max width=\textwidth}
\begin{threeparttable}
\begin{tabular}{ll|r|r|rr}
\toprule

& & \makecell{\textbf{Software (\sref{sec:eval-swe})}}
& \makecell{\textbf{Web (\sref{sec:eval_browsing})}}
& \multicolumn{2}{l}{\makecell{\textbf{Misc. (\sref{sec:eval_misc_assistance})}}} \\
\textbf{Agent}
& \textbf{Model}
& SWE-Bench Lite
& WebArena
& GPQA
& GAIA \\
\midrule

\rowcolor[RGB]{234, 234, 234} \multicolumn{6}{c}{\textit{Software Engineering Agents}} \\
SWE-Agent \citep{yang2024sweagent} & \texttt{gpt-4-1106-preview} & $18.0$ & $-$ & $-$ & $-$  \\
AutoCodeRover \citep{zhang2024autocoderover} & \texttt{gpt-4-0125-preview} & $19.0$ & $-$ & $-$ & $-$ \\
Aider \citep{aider} & \texttt{gpt-4o} \& \texttt{claude-3-opus} & $26.3$ & $-$ & $-$ & $-$ \\

Moatless Tools \citep{moatless_tool} & \texttt{claude-3.5-sonnet} & $26.7$ & $-$ & $-$ & $-$ \\
Agentless \citep{agentless} & \texttt{gpt-4o} & $27.3$ & $-$ & $-$ & $-$ \\

\midrule
\rowcolor[RGB]{234, 234, 234} \multicolumn{6}{c}{\textit{Web Browsing Agents}} \\

Lemur~\citep{xu2023lemur} & \texttt{Lemur-chat-70b} & $-$ & $5.3$ & $-$ & $-$ \\
\citet{patel2024large} & Trained 72B w/ synthetic data & $-$ & $9.4$ & $-$ & $-$ \\
AutoWebGLM~\citep{lai2024autowebglm} & Trained 7B w/ human/agent annotation & $-$ & $18.2$ & $-$ & $-$\\ 
Auto Eval \& Refine~\citep{pan2024autonomous} & GPT-4 + Reflexion w/ GPT-4V & $-$ & $20.2$ & $-$ & $-$ \\
\midrule

WebArena Agent~\citep{zhou2023webarena} & \texttt{gpt-4-turbo} & $-$ & $14.4$ &  $-$ & $-$ \\

\midrule
\rowcolor[RGB]{234, 234, 234} \multicolumn{6}{c}{\textit{Misc. Assistance Agents}} \\

\multirow{1}{*}{AutoGPT \citep{gravitasauto}} & \texttt{gpt-4-turbo} & $-$ & $-$ & $-$ & $13.2$ \\

\cmidrule{1-6}
\multirow{3}{*}{\makecell[l]{Few-shot Prompting \\
+ Chain-of-Thought \citep{rein2023gpqa}}}
& \texttt{Llama-2-70b-chat} & $-$ & $-$ & $28.1$ & $-$ \\
& \texttt{gpt-3.5-turbo-16k} & $-$ & $-$ & $29.6$ & $-$ \\
& \texttt{gpt-4} & & $-$ & $38.8$ & $-$ \\

\midrule
\rowcolor[RGB]{234, 234, 234} \multicolumn{6}{c}{\textbf{\projname Agents}} \\

\multirow{3}{*}{CodeActAgent \texttt{v1.8}} 
& \texttt{gpt-4o-mini-2024-07-18} & $6.3$ & $8.3$ & $-$ & $-$ \\
& \texttt{gpt-4o-2024-05-13} & $22.0$ & $14.5$ & $^{*}53.1$ & $-$ \\
& \texttt{claude-3-5-sonnet} & $26.0$ & $15.3$ & $52.0$ & $-$ \\

\cmidrule{1-6}
GPTSwarm \texttt{v1.0} & \texttt{gpt-4o-2024-05-13} & $-$ & $-$ & $-$ & $32.1$ \\

\bottomrule

\end{tabular}
{
\small
\begin{tablenotes}
\item[*] Numbers are reported from CodeActAgent \texttt{v1.5}.
\end{tablenotes}
}
\end{threeparttable}
\vspace{-2cm}
\end{adjustbox}
\label{tab:overall_perf}
\end{table*}
}

\subsection{Software Engineering}
\label{sec:eval-swe}

Next, we report results specifically for software engineering benchmarks in \tref{tab:swe_bench}.

\textbf{SWE-Bench} \citep{jimenez2024swebench} is designed to assess agents' abilities in solving real-world GitHub issues, such as bug reports or feature requests.
The agent interacts with the repository and attempts to fix the issue provided through file editing and code execution.
The agent-modified code repository is tested against a test suite incorporating new tests added from human developers' fixes for the same issue.
Each test instance accompanies a piece of ``hint text'' that consists of natural language suggestions for how to solve the problem. Throughout this paper, we report all results \textit{without using hint text}.
A canonical subset, SWE-bench Lite, is created to facilitate accessible and efficient testing. We default to use this subset for testing for cost-saving consideration.\footnote{Running the complete set of 2294 instances costs \$6.9k, using a conservative estimate of \$3 per instance.}
\textbf{Result.} As shown in \tref{tab:swe_bench}, our most recent version of CodeActAgent v1.8, using \texttt{claude-3.5-sonnet}, achieves a competitive resolve rate of $26\%$ compared to other open-source SWE specialists.

{
\begin{table*}[!ht]
\centering
\vspace{-0.5cm}
\caption{\projname Software Engineering evaluation results (\sref{sec:eval-swe}).}
\vspace{-0.2cm}
\begin{adjustbox}{max width=\textwidth}
\begin{threeparttable}
\begin{tabular}{l|l|rr}
\toprule
\textbf{Agent}
& \textbf{Model}
& \textbf{Success Rate} (\%)
& \textbf{\$ Avg. Cost} \\
\toprule
\rowcolor[RGB]{234, 234, 234} \multicolumn{4}{c}{\textbf{SWE-Bench Lite} \citep{jimenez2024swebench}, 300 instances, \textit{w/o Hint}} \\

SWE-Agent \citep{yang2024sweagent} & \texttt{gpt-4-1106-preview} & $18.0$ & $1.67$ \\
AutoCodeRover \citep{zhang2024autocoderover} & \texttt{gpt-4-0125-preview} & $19.0$ & $-$ \\
Aider \citep{aider} & \texttt{gpt-4o} \& \texttt{claude-3-opus} & $26.3$ & $-$ \\

\midrule
\multirow{3}{*}{OH CodeActAgent v1.8} 
& \texttt{gpt-4o-mini-2024-07-18} & $7.0$ & $0.01$ \\
& \texttt{gpt-4o-2024-05-13} & $22.0$ & $1.72$ \\
& \texttt{claude-3-5-sonnet@20240620} & $26.0$ & $1.10$ \\

\toprule
\rowcolor[RGB]{234, 234, 234} \multicolumn{4}{c}{\textbf{HumanEvalFix} \citep{muennighoff2024octopack}, 164 instances} \\
\multirow{4}{*}{Prompting, 0-shot}
& \texttt{BLOOMZ-176B} & $16.6$ &  $-$\\
& \texttt{OctoCoder-15B} & $30.4$ &  $-$\\
& \texttt{DeepSeekCoder-33B-Instruct} & $47.5$ & $-$ \\
& \texttt{StarCoder2-15B} & $48.6$ & $-$ \\
\midrule
SWE-agent, 1-shot~\citep{yang2024sweagent} & \texttt{gpt-4-turbo} & $87.7$ & $-$ \\
\midrule
\multirow{2}{*}{OH CodeActAgent v1.5, Generalist, 0-shot.}
& \texttt{gpt-3.5-turbo-16k-0613} & $20.1$ &  $0.11$\\
& \texttt{gpt-4o-2024-05-13} & $79.3$ & $0.14$ \\

\toprule
\rowcolor[RGB]{234, 234, 234} \multicolumn{4}{c}{\textbf{BIRD} \citep{bird}, 300 instances} \\
\multirow{2}{*}{Prompting, 0-shot} & \texttt{CodeLlama-7B-Instruct} & $18.3$ & - \\
& \texttt{CodeQwen-7B-Chat} & $31.3$ & - \\
\midrule
\multirow{2}{*}{OH CodeActAgent v1.5} 
& \texttt{gpt-4-1106-preview} & $42.7$ & $0.19$ \\
& \texttt{gpt-4o-2024-05-13} & $47.3$ & $0.11$ \\

\toprule
\rowcolor[RGB]{234, 234, 234} \multicolumn{4}{c}{\textbf{ML-Bench} \citep{liu2024mlbench}, 68 instances} \\
\multirow{2}{*}{prompting + BM25, 0-shot} & \texttt{gpt-3.5-turbo} & $11.0$ & - \\
& \texttt{gpt-4-1106-preview} & $22.1$ & - \\
& \texttt{gpt-4o-2024-05-13} & $26.2$ & - \\
\midrule
SWE-Agent \citep{yang2024sweagent} & \texttt{gpt-4-1106-preview} & $42.6$ & $1.91$ \\
Aider \citep{aider} & \texttt{gpt-4o} & $64.4$ & - \\
\midrule
\multirow{3}{*}{OH CodeActAgent v1.5} 
& \texttt{gpt-4o-2024-05-13} & $76.5$ & $0.25$ \\
& \texttt{gpt-4-1106-preview} & $58.8$ & $1.22$ \\
& \texttt{gpt-3.5-turbo-16k-0613} & $13.2$ & $0.12$\\

\toprule
\rowcolor[RGB]{234, 234, 234} 
\multicolumn{4}{c}{\textbf{BioCoder (Python)} \citep{liu2024mlbench}, 157 instances} \\
\multirow{2}{*}{prompting, 0-shot} & \texttt{gpt-3.5-turbo} & $11.0$ & - \\
& \texttt{gpt-4-1106-preview} & $12.7$ & - \\
\midrule
\multirow{1}{*}{OH CodeActAgent v1.5} 
& \texttt{gpt-4o-2024-05-13} & $27.5$ & $0.13$ \\

\toprule
\rowcolor[RGB]{234, 234, 234} \multicolumn{4}{c}{\textbf{Gorilla APIBench} \citep{patil2023gorilla}, 1775 instances } \\
\multirow{3}{*}{Prompting, 0-shot} 
& \texttt{claude-v1} & $8.7$ & -\\
& \texttt{gpt-4-0314} & $21.2$ & - \\
& \texttt{gpt-3.5-turbo-0301} & $29.7$ & - \\
\midrule
Gorilla, finetuned for API calls, 0-shot \citep{patil2023gorilla, touvron2023llama} & \texttt{llama-7b} & 75.0 & - \\
\midrule
\multirow{2}{*}{OH CodeActAgent v1.5} 
& \texttt{gpt-3.5-turbo-0125} & $21.6$ &  $0.002$ \\
& \texttt{gpt-4o-2024-05-13} & $36.4$ & $0.04$ \\

\toprule
\rowcolor[RGB]{234, 234, 234} \multicolumn{4}{c}{\textbf{ToolQA} \citep{zhuang2024toolqa}, 800 instances } \\
\multirow{3}{*}{Prompting, 0-shot} 
& \texttt{ChatGPT + CoT} & $5.1$ & - \\
& \texttt{ChatGPT} & $5.6$ & - \\
& \texttt{Chameleon} & $10.6$ & - \\
\midrule
\multirow{2}{*}{ReAct, 0-shot \citep{yao2023react,openai2024chatgpt}} 
& \texttt{gpt-3.5-turbo} & $36.8$ & - \\
& \texttt{gpt-3} & $43.1$ & - \\
\midrule
\multirow{2}{*}{OH CodeActAgent v1.5} 
& \texttt{gpt-3.5-turbo-0125} & $2.3$ & $0.03$ \\
& \texttt{gpt-4o-2024-05-13} & $47.2$ & $0.91$ \\

\bottomrule
\end{tabular}
\end{threeparttable}
\end{adjustbox}
\label{tab:swe_bench}
\vspace{-0.6cm}
\end{table*}
}

\subsubsection{HumanEvalFix}

\textbf{HumanEvalFix} \citep{muennighoff2024octopack} tasks agents to fix a bug in a provided function with the help of provided test cases. The bugs are created to ensure one or more test cases fail. We focus on the Python subset of the benchmark and allow models to solve the bugs by self-debug over multiple turns, incorporating feedback from test execution. We follow the setup from \citet{muennighoff2024octopack} using pass@k~\citep{chen2021evaluating}.
\textbf{Results.} In \tref{tab:swe_bench}, \projname CodeActAgent successfully fixes $79.3\%$ of bugs in the Python split. This is significantly better than all non-agentic approaches, almost doubling the performance of \texttt{StarCoder2-15B}~\citep{lozhkov2024starcoder2stackv2,li2023starcodersourceyou}.
While SWE-Agent achieves $87.7\%$, \citet{yang2024sweagent} provides the model a full demonstration of a successful sample trajectory fixing one of the bugs in the test dataset (``1-shot''), whereas our evaluation of \projname is 0-shot.
As HumanEvalFix has been created by humans and all bugs carefully validated, achieving $100\%$ on this benchmark is entirely feasible, which we seek to do in future iterations of \projname.

\textbf{ML-Bench} \citep{liu2024mlbench} 
evaluates agents' ability to solve machine learning tasks across 18 GitHub repositories.
The benchmark comprises 9,641 tasks spanning 169 diverse ML problems, requiring agents to generate bash scripts or Python code in response to user instructions.
In the sandbox environment, agents can iteratively execute commands and receive feedback, allowing them to understand the repository context and fulfill user requirements progressively.
Following the setup from the original paper, we perform agent evaluation on the quarter subset of ML-Bench.

\textbf{Gorilla APIBench} \citep{patil2023gorilla} evaluates agents' abilities to use APIs. it incorporates tasks on TorchHub, TensorHub, and HuggingFace. During the evaluation, models are given a question related to API usage, such as "\textit{identify an API capable of converting spoken language in a recording to text}." Correctness is evaluated based on whether the model’s API call is in the correct domain.

\textbf{ToolQA} \citep{zhuang2024toolqa} evaluates agents' abilities to use external tools. This benchmark includes tasks on various topics like flight status, coffee price, Yelp data, and Airbnb data, requiring the use of various tools such as text tools, database tools, math tools, graph tools, code tools, and system tools.
It features two levels: easy and hard. Easy questions focus more on single-tool usage, while hard questions emphasize reasoning. We adopt the easy subset for evaluation.

\textbf{BioCoder} \citep{tang2023biocoder} is a repository-level code generation benchmark that evaluates agents' performance on bioinformatics-related tasks, specifically the ability to retrieve and accurately utilize context. The original prompts contain the relevant context of the code; however, in this study, we have removed them to demonstrate the capability of \projname to perform context retrieval, self-debugging, and reasoning in multi-turn interactions. BioCoder consists of 157 Python and 50 Java functions, each targeting a specific area in bioinformatics, such as proteomics, genomics, and other specialized domains. The benchmark targets real-world code by generating code in existing repositories where the relevant code has been masked out.

\textbf{BIRD} \citep{bird} is a benchmark for text-to-SQL tasks (\ie, translate natural language into executable SQL) aimed at realistic and large-scale database environments.
We select 300 samples from the dev set to integrate into \projname and evaluate on execution accuracy. 
Additionally, we extend the setting by allowing the agent to engage in multi-turn interactions to arrive at the final SQL query, enabling it to correct historical results by observing the results of SQL execution.

\subsection{Web Browsing}
\label{sec:eval_browsing}

We report evaluation results for web browsing benchmarks in \tref{tab:web_browsing}.

\textbf{WebArena}~\citep{zhou2023webarena} is a self-hostable, execution-based web agent benchmark that allows agents to freely choose which path to take in completing their given tasks. 
WebArena comprises 812 human-curated task instructions across various domains, including shopping, forums, developer platforms, and content management systems. 
\textbf{Results.} From \tref{tab:web_browsing}, we can see that our BrowsingAgent achieves competitive performance among agents that use LLMs with domain-general prompting techniques.

\textbf{MiniWoB++}~\citep{liu2018reinforcement} is an interactive web benchmark, with built-in reward functions.
The tasks are synthetically initialized on 125 different minimalist web interfaces.
Unlike WebArena, tasks are easier without page changes, require fewer steps, and provide low-level step-by-step task directions.
Note that it contains a portion of environments that require vision capability to tackle successfully, and many existing work choose to focus only on a subset of the tasks~\citep{kim2024language,li2023zero,shaw2023pixels}. Still, we report the performance on the full set and only include baselines that are evaluated on the full set.

{
\begin{table*}[t]
\centering

\vspace{-0.5cm}
\caption{\projname Web Browsing Evaluation Results (\sref{sec:eval_browsing}).}
\vspace{-0.2cm}
\begin{adjustbox}{max width=\textwidth}
\begin{tabular}{l|l|rr}
\toprule
\textbf{Agent} & \textbf{Model} & \textbf{Success Rate}  (\%) & \textbf{\$ Avg. Cost} \\
\toprule

\rowcolor[RGB]{234, 234, 234} \multicolumn{4}{c}{\textbf{WebArena} \citep{zhou2023webarena}, 812 instances} \\
Lemur~\citep{xu2023lemur} & \texttt{Lemur-chat-70b} & $5.3$ & $-$ \\
\citet{patel2024large} & Trained 72B with self-improvement synthetic data & $9.4$ & $-$ \\
AutoWebGLM~\citep{lai2024autowebglm} & Trained 7B with human/agent hybrid annotation & $18.2$ & $-$\\ 
Auto Eval \& Refine~\citep{pan2024autonomous} & GPT-4 + Reflexion w/ GPT-4V reward model & $20.2$ & $-$\\
\midrule
\multirow{2}{*}{WebArena Agent~\citep{zhou2023webarena}} 
& \texttt{gpt-3.5-turbo} & $6.2$ & $-$  \\
& \texttt{gpt-4-turbo} & $14.4$ &  $-$  \\
\midrule
\multirow{3}{*}{OH BrowsingAgent v1.0} 
& \texttt{gpt-4o-mini-2024-07-18} & $8.5$ &$0.01$  \\
& \texttt{gpt-4o-2024-05-13} & $14.8$ &$0.15$  \\
& \texttt{claude-3-5-sonnet-20240620} & $15.5$ &$0.10$ \\

\midrule

\multirow{3}{*}{\makecell[l]{OH CodeActAgent v1.8\\ {\small via \textbf{delegation} to BrowsingAgent v1.0}}}
& \texttt{gpt-4o-mini-2024-07-18} & $8.3$ & $-$ \\
& \texttt{gpt-4o-2024-05-13} & $14.5$ & $-$ \\
& \texttt{claude-3-5-sonnet-20240620} & $15.3$ &$-$ \\

\toprule
\rowcolor[RGB]{234, 234, 234} \multicolumn{4}{c}{\textbf{MiniWoB++} \citep{liu2018reinforcement}, 125 environments} \\
Workflow Guided Exploration~\citep{liu2018reinforcement} & Trained specialist model with environment exploration & $34.6$ & $-$\\ 
CC-NET~\citep{humphreys2022data} & Trained specialist model with RL and human annotated BC & $91.1$ & $-$\\ 
\midrule
\multirow{2}{*}{OH BrowsingAgent v1.0} 
& \texttt{gpt-3.5-turbo-0125} & $27.2$ & $0.01$\\
& \texttt{gpt-4o-2024-05-13} & $40.8$ & $0.05$ \\

\midrule

\makecell[l]{OH CodeActAgent v1.8\\ {\small via \textbf{delegation} to BrowsingAgent v1.0}}
& \texttt{gpt-4o-2024-05-13} & $39.8$ & $-$ \\

\bottomrule
\end{tabular}
\end{adjustbox}
\vspace{-0.5cm}
\label{tab:web_browsing}
\end{table*}
}

\subsection{Miscellaneous Assistance}
\label{sec:eval_misc_assistance}

Results for miscellaneous assistance benchmarks are reported in \tref{tab:misc_assistance}. 

\textbf{GAIA}~\citep{gaia} evaluates agents' general task-solving skills, covering different real-world scenarios.
It requires various agent capabilities, including reasoning, multi-modal understanding, web browsing, and coding. 
GAIA consists of 466 curated tasks across three levels.
Setting up GAIA is traditionally challenging due to the complexity of integrating various tools with the agent, but 
\projname's infrastructure (\eg, runtime \sref{sec:agent-runtime}, tools \sref{sec:agent-skills}) simplifies the integration significantly. 

\textbf{GPQA} \citep{rein2023gpqa} evaluates agents' ability for coordinated tool use when solving challenging graduate-level problems.
Tool use (\eg, python) and web search are often useful to assist agents in answering these questions since they provide accurate calculations that LLMs are often incapable of and access to information outside of the LLM's parametric knowledge base.

{
\begin{table*}[t]
\centering
\caption{\projname miscellaneous assistance evaluation results (\sref{sec:eval_misc_assistance}).}
\vspace{-0.2cm}
\begin{adjustbox}{max width=\textwidth}
\begin{tabular}{l|l|rr}
\toprule
\textbf{Agent} & \textbf{Model} & \textbf{Success Rate} (\%) & \textbf{\$ Avg. Cost} \\
\toprule

\rowcolor[RGB]{234, 234, 234} \multicolumn{4}{c}{\textbf{GAIA} \citep{gaia}, L1 validation set, 53 instances} \\

\multirow{1}{*}{AutoGPT \citep{gravitasauto}} & \texttt{gpt-4-turbo} & $13.2$ & $-$ \\
\midrule
\multirow{2}{*}{OH GPTSwarm v1.0} 
& \texttt{gpt-4-0125-preview} & $30.2$ & $0.110$ \\
& \texttt{gpt-4o-2024-05-13} & $32.1$ & $0.050$ \\

\toprule
\rowcolor[RGB]{234, 234, 234} \multicolumn{4}{c}{\textbf{GPQA} 
\citep{rein2023gpqa}, diamond set, 198 instances (refer to \sref{sec:gpqa-additional}, \tref{tab:gpqa} for other subsets)} \\

\multirow{2}{*}{Human \citep{rein2023gpqa}} 
& Expert human & $81.3$ & $-$ \\
& Non-expert human & $21.9$ & $-$ \\

\midrule

\multirow{2}{*}{Few-shot Prompting + Chain-of-Thought \citep{rein2023gpqa}} 
& \texttt{gpt-3.5-turbo-16k} & $29.6$ & $-$ \\
& \texttt{gpt-4} & $38.8$ & $-$ \\

\midrule
\multirow{1}{*}{OH CodeActAgent v1.8} 
& \texttt{claude-3-5-sonnet-20240620} & $52.0$ & $0.065$ \\

\toprule
\rowcolor[RGB]{234, 234, 234} \multicolumn{4}{c}{\textbf{AgentBench} \citep{liu2023agentbench}, OS (bash) subset, 144 instances} \\

\multirow{2}{*}{AgentBench Baseline Agent \citep{liu2023agentbench}}
& \texttt{gpt-4} & $42.4$ & $-$ \\
& \texttt{gpt-3.5-turbo} & $32.6$ & $-$ \\
\midrule
\multirow{2}{*}{OH CodeActAgent v1.5}
& \texttt{gpt-4o-2024-05-13} & $57.6$ & $0.085$\\
& \texttt{gpt-3.5-turbo-0125} & $11.8$ & $0.006$\\

\toprule
\rowcolor[RGB]{234, 234, 234} \multicolumn{4}{c}{\textbf{MINT} \citep{wang2024mint}: \texttt{math} subset, 225 instances} \\

\multirow{1}{*}{MINT Baseline Agent}
& \texttt{gpt-4-0613} & $65.8$ & $-$ \\

\midrule

\multirow{2}{*}{OH CodeActAgent v1.5}
& \texttt{gpt-4o-2024-05-13} & $77.3$ & $0.070$ \\
& \texttt{gpt-3.5-turbo-16k-0613} & $33.8$ & $0.048$ \\

\toprule
\rowcolor[RGB]{234, 234, 234} \multicolumn{4}{c}{\textbf{MINT} \citep{wang2024mint}: \texttt{code} subset, 136 instances} \\

\multirow{1}{*}{MINT Baseline Agent}
& \texttt{gpt-4-0613} & $59.6$ & $-$ \\

\midrule

\multirow{2}{*}{OH CodeActAgent v1.5}
& \texttt{gpt-4o-2024-05-13} & $50.0$ & $0.087$ \\
& \texttt{gpt-3.5-turbo-16k-0613} & $5.2$ & $0.030$ \\

\toprule
\rowcolor[RGB]{234, 234, 234} \multicolumn{4}{c}{\textbf{ProofWriter} \citep{proofwriter}, 600 instances} \\
\multirow{1}{*}{Few-shot Prompting + Chain-of-Thought \citep{logic_lm}}  
& \texttt{gpt4} & $68.1$ & $-$ \\
\midrule
\multirow{1}{*}{Logic-LM \citep{logic_lm}}
& \texttt{gpt4 + symbolic solver} & $79.6$ & $-$ \\
\midrule
\multirow{1}{*}{OH CodeActAgent v1.5}
& \texttt{gpt-4o-2024-05-13} & $78.8$ & $-$ \\

\toprule
\rowcolor[RGB]{234, 234, 234} \multicolumn{4}{c}{\textbf{Entity Deduction Arena} \citep{zhang2023entity}, 200 instances} \\
\multirow{2}{*}{Zero-shot Prompting \citep{zhang2023entity}}
& \texttt{gpt-4-0314} & $40.0$ & $-$ \\
& \texttt{gpt-3.5-turbo-0613} & $27.0$ & $-$ \\
\midrule
\multirow{2}{*}{OH CodeActAgent v1.5}
& \texttt{gpt-4o-2024-05-13} & $38.0$ & $-$ \\
& \texttt{gpt-3.5-turbo-16k-0613} & $24.0$ & $-$ \\

\bottomrule
\end{tabular}
\end{adjustbox}
\label{tab:misc_assistance}
\vspace{-0.7cm}
\end{table*}
}

\textbf{AgentBench}~\citep{liu2023agentbench} evaluates agents' reasoning and decision-making abilities in a multi-turn, open-ended generation setting.
We selected the code-grounded operating system (OS) subset with 144 tasks.
Agents from \projname interact directly with the task-specific OS using bash commands in a multi-turn manner, combining interaction and reasoning to automate task completion.

\textbf{MINT} \citep{wang2024mint} is a benchmark designed to evaluate agents' ability to solve challenging tasks through \textit{multi-turn interactions} using \textit{tools} and \textit{natural language feedback} simulated by GPT-4. We use coding and math subsets used in \citet{yuan2024advancing}. We follow the original paper and allow the agent to interact with up to five iterations with two chances to propose solutions.

\textbf{ProofWriter} \citep{proofwriter} is a synthetic dataset created to assess deductive reasoning abilities of LLMs.
Same as Logic-LM \citep{logic_lm}, we focus on the most challenging subset, which contains 600 instances requiring 5-hop reasoning. To minimize the impact of potential errors in semantic parsing, we use the logical forms provided by Logic-LM.

\textbf{Entity Deduction Arena} (EDA) \citep{zhang2023entity} evaluates agents' ability to deduce unknown entities through strategic questioning, akin to the 20 Questions game. 
This benchmark tests the agent's state tracking, strategic planning, and inductive reasoning capabilities over multi-turn conversations. We evaluate two datasets ``Things'' and ``Celebrities'', each comprising 100 instances, and report the average success rate over these two datasets.

\looseness=-1
\section{Conclusion}

We introduce \projname, a community-driven platform that enables the development of agents that interact with the world through software interfaces.
By providing a powerful interaction mechanism, a safe sandboxed environment, essential agent skills, multi-agent collaboration capabilities, and a comprehensive evaluation framework, \projname accelerates research innovations and real-world applications of agentic AI systems.
Despite challenges in developing safe and reliable agents (\sref{sec:limitation-future-work}), we are excited about our vibrant community and look forward to \projname's continued evolution.

\bibliography{custom}
\bibliographystyle{iclr2025_conference}

\appendix
\newpage

\section*{Author Contributions}
This work was an open-source collaborative effort across multiple institutions. We employed a point-based system to determine contributions and award authorships, with technical contributions tracked and measured in units of pull requests (PRs)\footnote{For more details, please refer to \url{https://github.com/All-Hands-AI/OpenHands/pull/1917}.}. Xingyao Wang led the project, coordinating overall development and paper writing efforts. Detailed contributions were as follows:

\begin{itemize}[noitemsep,topsep=0pt,parsep=2pt,partopsep=0pt,leftmargin=18pt]

\item \textbf{Agent Development} (\sref{sec:agenthub}): Xingyao Wang led the implementation of CodeAct \cite{wang2024executable} and CodeActSWE agents. Frank F. Xu led the development of web browsing agents \cite{zhou2023webarena}. Mingchen Zhuge orchestrated the integration of the GPTSwarm agent \cite{zhuge2024language}. Robert Brennan and Boxuan Li lead the development of the Micro Agent.

\item \textbf{Architectural Development} (\fref{fig:arch}): Robert Brennan initiated the architecture design. Boxuan Li, Frank F. Xu, Xingyao Wang, Yufan Song, and Mingzhang Zheng further refined and expanded the architecture. Boxuan Li implemented the initial version of integration tests (\sref{sec:integration-tests}), maintained the agentskills library (\sref{sec:agent-skills}), managed configurations, and resolved resource leaks in evaluation. Frank F. Xu developed the web browsing environment (\sref{sec:browsergym-actions}) for both agent execution and evaluation and integrated it with both agent and front-end user interfaces. Xingyao Wang authored the initial code for the agentskills library and the Docker sandbox. Yufan Song implemented cost tracking for evaluation, while Mingzhang Zheng developed an image-agnostic docker sandbox for more stable SWE-Bench evaluation.

\item \textbf{Benchmarking, Integration, and Code Review}: Boxuan Li and Yufan Song led benchmark integration efforts, including coordination, evaluation, and code review. Yufan Song also helped track PR contributions. Graham Neubig, Xingyao Wang, Mingzhang Zheng, Robert Brennan, Hoang H. Tran, Frank F. Xu, Xiangru Tang, Fuqiang Li, and Yanjun Shao provided additional support in integration and code reviews.
Specific benchmark contributions included:
\begin{itemize}
    \item SWE-Bench \cite{jimenez2024swebench}: Bowen Li and Xingyao Wang
    \item WebArena \cite{zhou2023webarena} and MiniWob++ \cite{liu2018reinforcement}: Frank F. Xu
    \item GAIA \cite{gaia}: Jiayi Pan (integration) and Mingchen Zhuge (GPTSwarm evaluation)
    \item API-Bench \cite{patil2023gorilla} and ToolQA \cite{zhuang2024toolqa}: Yueqi Song
    \item HumanEvalFix \cite{muennighoff2024octopack}: Niklas Muennighoff and Xiangru Tang
    \item ProofWriter \cite{proofwriter}: Ren Ma
    \item MINT \cite{wang2024mint}: Hoang H. Tran
    \item AgentBench \cite{liu2023agentbench}: Fuqiang Li
    \item BIRD \cite{bird}: Binyuan Hui
    \item GPQA \cite{rein2023gpqa}: Jaskirat Singh
    \item BioCoder \cite{tang2023biocoder}: Xiangru Tang and Bill Qian
    \item ML-Bench \cite{liu2024mlbench}: Xiangru Tang and Yanjun Shao
    \item Entity-Deduction-Arena \cite{zhang2023entity}: Yizhe Zhang
\end{itemize}

\item \textbf{Advising}: Graham Neubig advised the project, providing guidance, resources, and substantial paper edits. Heng Ji and Hao Peng offered additional project advice and assisted with paper writing. Junyang Lin contributed advisory support and sponsored resources.
\end{itemize}

\section{Limitations and Future Work}
\label{sec:limitation-future-work}
We are excited about the foundations our vibrant community has laid in \projname and look forward to its continued evolution.
We identify several directions for future work:

\textbf{Enhanced multi-modality support.} While our current implementation already supports a wide range of file formats through predefined agent skills, we are interested in enabling multi-modality in a principled way through standard IPython and browser integration, such as viewing images and videos using vision-language model through a browser or processing XLSX files with code.

\textbf{Stronger agents.} Current agents still struggle with complex tasks, and we are interested in building better agents through both training and inference time techniques. 

\textbf{Agent editing improvements.} Current agent suffers a lot when editing long files, and we are interested in exploring different approaches to improve the file editing performance of agents.

\textbf{Web browsing improvements.} Due to the extensible nature of \projname, orthogonal components that could improve agents can be integrated easily. For example, thanks to \projname's extensible architecture, Auto Eval \& Refine~\cite{pan2024autonomous}, an agent retry-on-error strategy with Reflexion~\cite{shinn2024reflexion} prompts and task completion reward models, will be integrated as an optional component attached to our browsing agent.

\textbf{Automatic workflow generation.} Currently, OpenHands's workflow still requires a substantial handcrafted workload. We believe that graph-based frameworks such as GPTSwarm~\cite{zhuge2024language} and LangGraph~\cite{langchain2022} could serve as alternative solutions for building agents. Particularly in GPTSwarm, when agents are constructed using graphs, it becomes easier to incorporate various optimization methods (e.g., reinforcement learning, meta-prompting). OpenHands considers these methods to lay the groundwork for promising solutions in automatic workflow generation in future versions.

\section{Ethics Statement}
\label{sec:ethics}
Most AI agents today are still research artifacts and lack the ability to perform complex, long-horizon tasks in the real world reliably. However, as their performance continues to improve and they are increasingly deployed in real world, they have the potential to boost productivity while also posing security risks to society significantly.
\projname helps mitigate risks by: 
\begin{itemize}[noitemsep,topsep=0pt,parsep=2pt,partopsep=0pt,leftmargin=18pt]
\item[(1)] Enabling systematic evaluation of these agents, which can identify and address risks before they are widely deployed.
\item[(2)] Facilitating human-agent interaction rather than allowing agents to operate autonomously without oversight.
\item[(3)] More importantly, we hope \projname allows researchers worldwide to access the best suites of agents to conduct frontier safety research towards building safe and helpful agents.
\end{itemize}

\section{Related Work}
\label{sec:related_work}

The breakthroughs in large language models (LLMs) like ChatGPT~\cite{openai2024chatgpt} and GPT-4~\cite{openai2024gpt4} have significantly enhanced the capabilities of autonomous agents across various domains~\cite{ye2023proagent,tang2024medagents,park2023generative,cui2023chatlaw}. These advances have spurred a multitude of generalist agent proposals~\cite{gravitasauto,nakajima2023task,wu2023autogen} aimed at performing diverse user tasks and have gained attention from both developers and broader audiences. Notable works such as Auto-GPT~\cite{gravitasauto} harness LLMs for task completion by decomposing user goals into executable steps. Multi-agent collaboration systems leverage LLMs for elements like role-playing and task-solving capabilities~\cite{zhuge2023mindstorms,li2023camel,zhou2023agents,xagent2023}, with MetaGPT~\cite{hong2023metagpt} emphasizing standardized operating procedures, and AutoGen~\cite{wu2023autogen} providing a conversation framework for interactive systems. AGENTS~\cite{zhou2023agents} and AutoAgents~\cite{chen2024autoagents} offer new paradigms for customizable agent architecture, while XAgent~\cite{xagent2023} and GPTSwarm~\cite{zhuge2024language} introduce complex management systems and optimizable graphs, respectively, for enhanced agent operations.

This surge in agent development has led to specialized frameworks aimed at streamlining agent implementation. LangChain and LangGraph \cite{langchain2022} provide foundational building blocks with basic runtime support, while CrewAI \cite{crewai2024} focuses on orchestrating multi-agent communications. BrowserGym \cite{browsergym} specifically targets web browsing capabilities, and DSPy \cite{khattab2024dspy} emphasizes end-to-end prompt optimization. AutoGen~\cite{wu2023autogen} advances beyond basic frameworks by implementing Python and bash execution capabilities, though with stateless command execution, while frameworks like CrewAI offer sandboxed but limited code interpreter features.

Software development, a front-runner in applying LLM-based agents, has seen advancements in frameworks for facilitating the development processes~\cite{hong2023metagpt,qian2023communicative}. Innovations such as ChatDev~\cite{qian2023communicative} automate the software development lifecycle akin to the waterfall model, and AutoCodeRover~\cite{zhang2024autocoderover} addresses GitHub issues via code search and abstract syntax tree manipulation. AgentCoder~\cite{huang2024agentcoder} iteratively refines code generation with integrated testing and feedback, while SWE-Agent~\cite{yang2024sweagent} integrates LLMs for automated Github issue fixing, streamlining software engineering.

\section{Graphical User Interface}
\label{sec:ui}

Besides running from the command line, \projname features a rich graphical user interface that visualizes the agent's current actions (\eg, browsing the web, executing base commands or Python code, \etc) and allows for real-time feedback from the user.
Screenshots of the UI are shown in \fref{fig:ui}.
The user may interrupt the agent at any moment to provide additional feedback, comments, or instruction while the agent is working.
This user interface directly connects with the event streams (\sref{sec:agent-abstraction}) to control and visualize the agents and runtime, making it agent and runtime agnostic.

\section{Quality Control: Integration Tests for Agents}
\label{sec:integration-tests}

Integration tests \cite{DBLP:conf/icsm/LeungW90} have long been used by software developers to ensure software quality. Unlike large language models with simple input-output schema, agents are typically complex pieces of software where minor errors can be easily introduced during the development process and hurt final task performance.
While running a full suite evaluation (\sref{sec:evaluation}) is the ultimate measure of performance degradation, running them for \textit{every} code changes can be prohibitively slow and expensive.
\footnote{Running a SWE-Bench Lite \cite{jimenez2024swebench} evaluation with \texttt{gpt-4o} costs around 600 USD.}.
In \projname, we pioneer an end-to-end agent test framework that tests prompt regression, actions, and sandbox environments.
It combines integration testing from software engineering and foundation model mocking for deterministic behavior to prevent the accidental introduction of bugs during agent development.

\noindent \textbf{Defining an integration test.}
The integration test framework for OpenHands is structured to validate end-to-end functionality by automating task execution and result verification. Developers define tasks and expected results; for instance, a task might involve correcting typos in a document named "bad.txt". Upon task execution through OpenHands, outputs are compared against a predefined "gold file" to ensure accuracy.

\noindent \textbf{Mocking LLM for deterministic behavior.}
Addressing the challenge of non-determinism in large language models (LLMs) and the associated high costs, the framework intercepts all LLM calls and supplies predefined responses based on exact prompt matches. This method not only ensures consistency in test outcomes but also reduces operational costs by minimizing the reliance on real LLMs.

\noindent \textbf{Regenerate LLM responses on breaking changes.}
Prompt-response pairs are managed through a script that generates and stores these pairs when new tests are introduced or existing prompts are modified. For routine tests, the framework attempts to reuse existing LLM responses by slightly adjusting the prompts. Substantial changes that affect task handling require regeneration of these pairs using real LLMs.

\noindent \textbf{Benefits of integration tests.}
The framework offers several advantages, including 1) Prompt regression testing: Stored prompt-response pairs facilitate change tracking and provide a reference for new team members to understand LLM interactions, 2) Multi-platform support: Tests are automatically scheduled for every pull request and commit on the main branch, running across multiple platforms, environments, and agents, including Linux and Mac, and in local, SSH, and exec sandboxes, and 3) Comprehensive error detection: It captures errors in prompt generation, message passing, and sandbox execution, thereby maintaining a high test coverage.

\section{How \projname Runtime work}
\label{sec:runtime-details}

\subsection{Workflow}
\begin{figure}
\centering
\vspace{-0.2cm}
\includegraphics[height=0.8\textwidth]{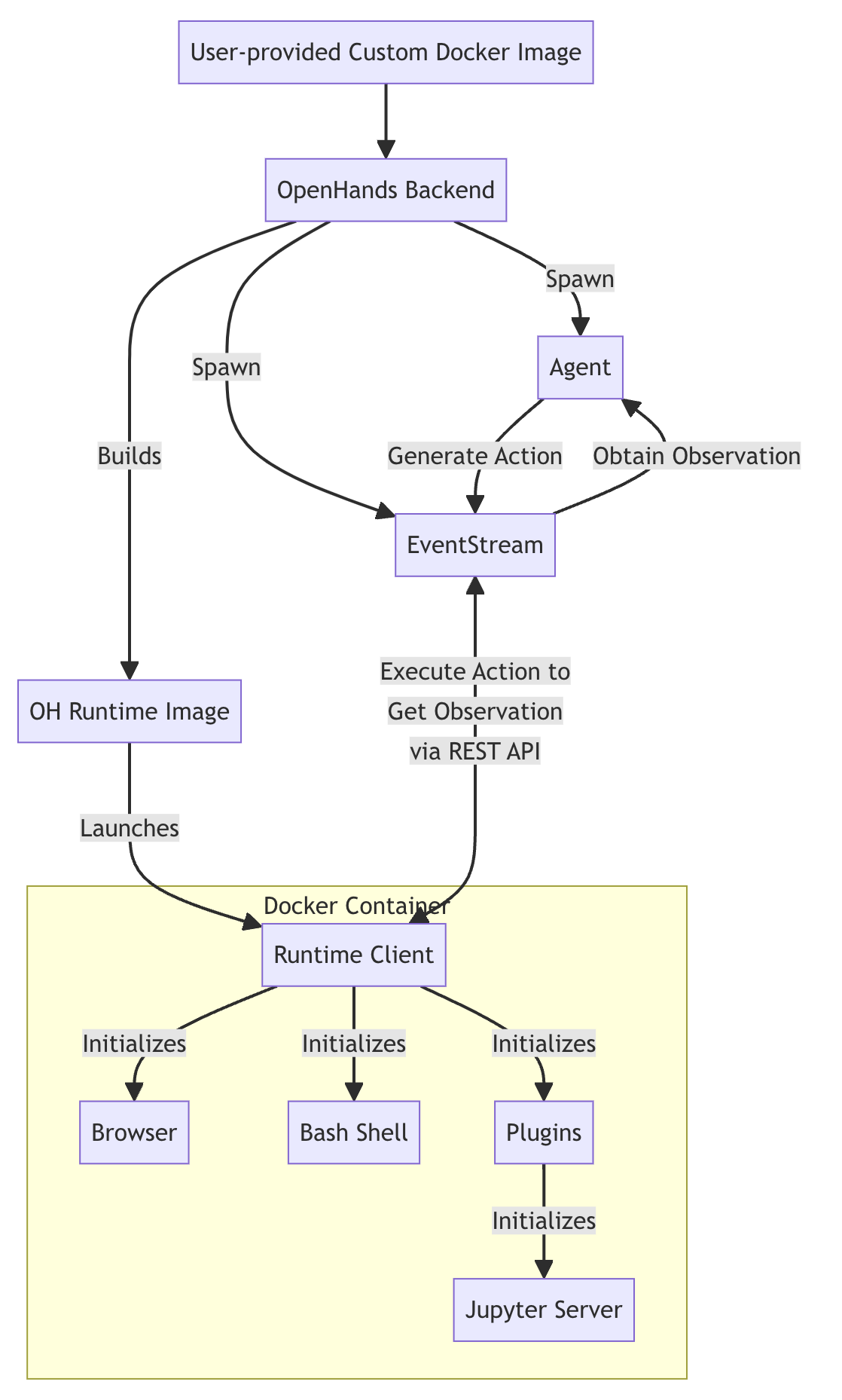}
\vspace{-0.2cm}
\caption{\projname runtime workflow.}
\label{fig:runtime_workflow}
\vskip -0.1in
\end{figure}

The OpenHands Runtime system uses a client-server architecture implemented with Docker containers. See \fref{fig:runtime_workflow} for an overview of how it works.

\begin{itemize}[noitemsep,topsep=0pt,parsep=2pt,partopsep=0pt,leftmargin=18pt]
\item[(1)] \textbf{User Input}: The user provides a custom base Docker image.

\item[(2)] \textbf{Image Building}: OpenHands builds a new Docker image (the "OH runtime image") based on the user-provided image. This new image includes OpenHands-specific code, primarily the "runtime client" (i.e., runtime API server described in \sref{sec:agent-runtime}).
\item[(3)] \textbf{Container Launch}: When OpenHands starts, it launches a Docker container using the OH runtime image.
\item[(4)] \textbf{Communication}: The OpenHands backend (\texttt{runtime.py}) communicates with the runtime client over RESTful API, sending actions and receiving observations
\item[(5)] \textbf{Action Execution}: The runtime client receives actions from the backend, executes them in the sandboxed environment, and sends back observations

\item[(6)] \textbf{Observation Return}: The client sends execution results back to the OpenHands backend event stream as observations.

\end{itemize}

The role of the client:
\begin{itemize}
    \item It acts as an intermediary between the OpenHands backend and the sandboxed environment
    \item It executes various types of actions (shell commands, file operations, Python code, etc.) safely within the container
    \item It manages the state of the sandboxed environment, including the current working directory and loaded plugins
    \item It formats and returns observations to the backend, ensuring a consistent interface for processing results
\end{itemize}

\subsection{How OpenHands builds and maintains runtime images}

OpenHands' approach to building and managing runtime images ensures efficiency, consistency, and flexibility in creating and maintaining Docker images for both production and development environments.

\subsubsection{Image Tagging System}

OpenHands uses a dual-tagging system for its runtime images to balance reproducibility with flexibility:

\begin{itemize}[noitemsep,topsep=0pt,parsep=2pt,partopsep=0pt,leftmargin=18pt]
\item[(1)] Hash-based tag: \texttt{\{target\_image\_repo\}:\{target\_image\_hash\_tag\}}.
   Example: \texttt{runtime:abc123def456}
    \begin{itemize}
        \item This tag is based on the MD5 hash of the Docker build folder, which includes the source code (of runtime client and related dependencies) and Dockerfile
        \item Identical hash tags guarantee that the images were built with exactly the same source code and Dockerfile
        \item This ensures reproducibility; the same hash always means the same image contents
    \end{itemize}

\item[(2)] Generic tag: \texttt{\{target\_image\_repo\}:\{target\_image\_tag\}}.
   Example: \texttt{runtime:oh\_v0.9.3\_ubuntu\_tag\_22.04}

   \begin{itemize}
       \item This tag follows the format: \texttt{runtime:oh\_v\{VERSION\}\_\{BASE\_IMAGE\}\_tag\_\{IMAGE\_TAG\}}
       \item It represents the latest build for a particular base image and OpenHands version combination
       \item This tag is updated whenever a new image is built from the same base image, even if the source code changes
   \end{itemize}
\end{itemize}

The hash-based tag ensures reproducibility, while the generic tag provides a stable reference to the latest version of a particular configuration. This dual-tagging approach allows OpenHands to efficiently manage both development and production environments.

\subsubsection{Build Process}

\begin{itemize}[noitemsep,topsep=0pt,parsep=2pt,partopsep=0pt,leftmargin=18pt]
\item[(1)] \textbf{Image Naming Convention}:
    \begin{itemize}
    \item Hash-based tag: \texttt{{target\_image\_repo}:{target\_image\_hash\_tag}}.
     Example: \texttt{runtime:abc123def456}
     \item Generic tag: \texttt{{target\_image\_repo}:{target\_image\_tag}}.
     Example: \texttt{runtime:oh\_v0.9.3\_ubuntu\_tag\_22.04}
    \end{itemize}

\item[(2)] \textbf{Build Process}:
   \begin{itemize}
   \item[a.] Convert the base image name to an OH runtime image name
      Example: \texttt{ubuntu:22.04} -> \texttt{runtime:oh\_v0.9.3\_ubuntu\_tag\_22.04}
   \item[b.] Generate a build context (Dockerfile and OpenHands source code) and calculate its hash
   \item[c.] Check for an existing image with the calculated hash
   \item[d.] If not found, check for a recent compatible image to use as a base
   \item[e.] If no compatible image exists, build from scratch using the original base image
   \item[f.] Tag the new image with both hash-based and generic tags
   \end{itemize}

\item[(3)] \textbf{Image Reuse and Rebuilding Logic}:
   The system follows these steps to determine whether to build a new image or use an existing one from a user-provided (base) image (e.g., \texttt{ubuntu:22.04}):
   \begin{itemize}
       \item[a.] If an image exists with the same hash (e.g., \texttt{runtime:abc123def456}), it will be reused as is
       \item[b.] If the exact hash is not found, the system will try to rebuild using the latest generic image (e.g., \texttt{runtime:oh\_v0.9.3\_ubuntu\_tag\_22.04}) as a base. This saves time by leveraging existing dependencies
       \item[c.] If neither the hash-tagged nor the generic-tagged image is found, the system will build the image completely from scratch
   \end{itemize}

\end{itemize}

\textbf{Caching and Efficiency.} The system attempts to reuse existing images when possible to save build time. If an exact match (by hash) is found, it's used without rebuilding. If a compatible image is found, it's used as a base for rebuilding, saving time on dependency installation.

\begin{figure}
\centering
\vspace{-0.2cm}
\includegraphics[width=0.5\textwidth]{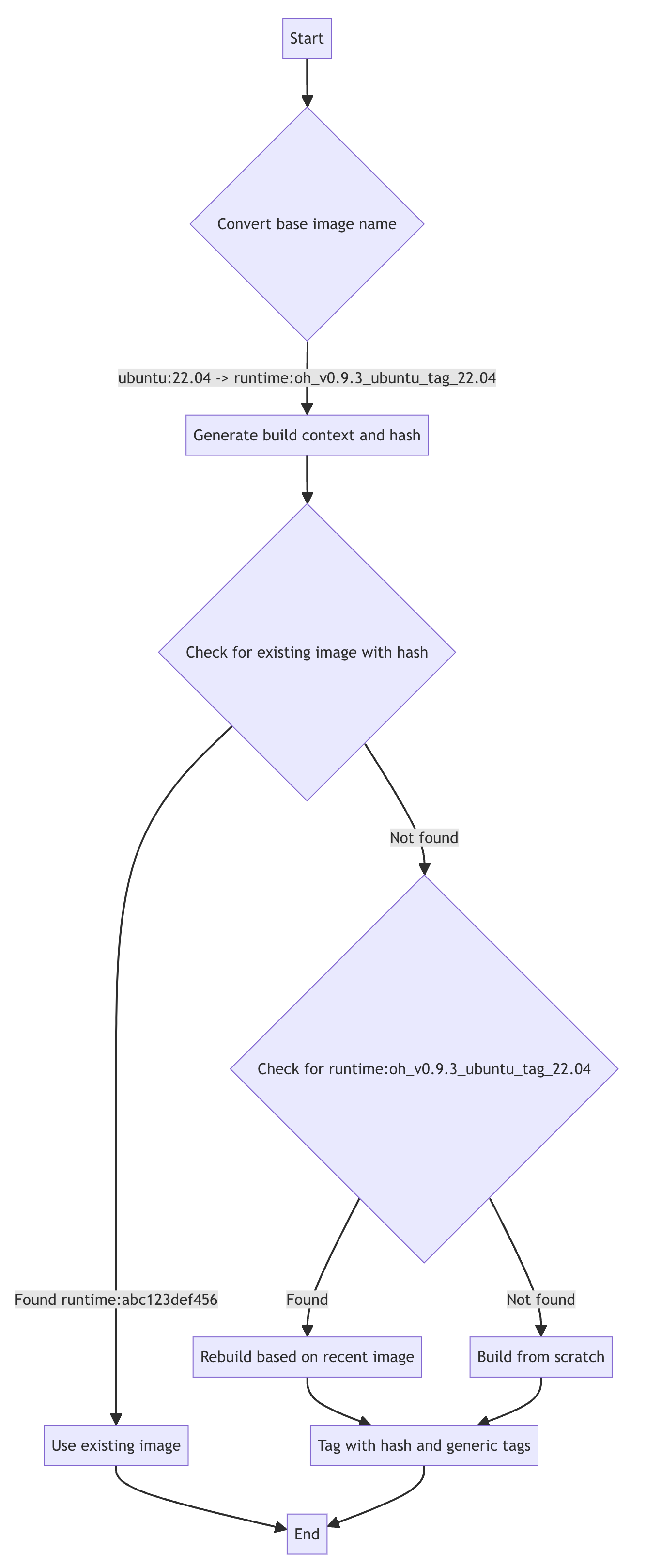}
\vspace{-0.2cm}
\caption{\projname Runtime Image Build Workflow.}
\label{fig:runtime-image-build-workflow}
\vskip -0.1in
\end{figure}

A flowchart illustrating the build process is shown in \fref{fig:runtime-image-build-workflow}

\section{Additional Results For GPQA Benchmark}
\label{sec:gpqa-additional}

We showcase more detailed results, including performance on other subsets for GPQA benchmark in~\tref{tab:gpqa}.

\begin{table}[ht]
    \centering
    \small
    \caption{\emph{Full Evaluation Results on the GPQA Benchmark \cite{rein2023gpqa}} (\sref{sec:eval_misc_assistance}).
    }
\begin{adjustbox}{max width=\textwidth}
    \begin{tabular}{lcccc}
        \toprule
        \multirow{2}{*}{\textbf{Evaluation Method and Model}}  & \multicolumn{3}{c}{\textbf{Accuracy by subset (\%)}} & \multirow{2}{*}{\textbf{Avg Cost (\$)}} \\[0.5ex]
        
         & {\emph{Diamond Set}} & {\emph{Main Set}} & {\emph{Extended Set}} & \\ 
         \midrule
         \rowcolor{Gray} Expert Human Validators & 81.2 & 72.5 & 65.4 & N/A \\
        \rowcolor{Gray} Non-Expert Human Validators & 21.9 & 30.5 & 33.9 & N/A \\
        \midrule
        Few-Shot CoT Llama-2-70B-chat & 28.1 & 29.1 & 30.4 & \cellcolor{Gray}N/A \\ 
        Few-Shot CoT GPT-3.5-turbo-16k & 29.6 & 28.0 & 28.2 & \cellcolor{Gray}N/A \\ 
        Few-Shot CoT GPT-4 & 38.8 & 39.7 & 38.7 & \cellcolor{Gray}N/A \\ 
        GPT-4 with search (backoff to CoT on abstention) &  38.8 &  41.0 & 39.4 & \cellcolor{Gray}N/A \\
        \midrule
        OpenHands + CodeActAgent v1.5 + GPT3.5-turbo & 27.9 & 23.4 & 26.1 & 0.012 \\
        OpenHands + CodeActAgent v1.5 + GPT4-turbo & 51.8 & 47.4 & 42.4 & 0.501 \\
        OpenHands + CodeActAgent v1.5 + GPT4o & \textbf{53.1} &  \textbf{49.3} & \textbf{52.8} & 0.054 \\
        \bottomrule
    \end{tabular}
    \label{tab:gpqa}
\end{adjustbox}
\end{table}

\section{In-context Demonstration for CodeActSWEAgent}
\label{sec:icl-codeactswe}

The prompt is re-adopted from the SWE-agent's released trajectory (\url{https://github.com/princeton-nlp/SWE-agent/tree/main/trajectories/demonstrations}). The prompt can be found at \url{https://github.com/All-Hands-AI/OpenHands/blob/main/agenthub/codeact_swe_agent/prompt.py}.

\section{Supported AgentSkills}
\label{sec:supported-agentskills}

As of \projname v0.6, we support the following list of skills.
Please refer to the source code for the most up-to-date list of skills: \url{https://github.com/All-Hands-AI/OpenHands/blob/main/OpenHands/runtime/plugins/agent_skills/agentskills.py}

\begin{minted}[breaklines]{python}
def open_file(path: str, line_number: Optional[int] = None) -> None:
    """
    Opens the file at the given path in the editor. If line_number is provided, the window will be moved to include that line.

    Args:
        path: str: The path to the file to open.
        line_number: Optional[int]: The line number to move to.
    """
    pass


def goto_line(line_number: int) -> None:
    """
    Moves the window to show the specified line number.

    Args:
        line_number: int: The line number to move to.
    """
    pass

def scroll_down() -> None:
    """Moves the window down by 100 lines.

    Args:
        None
    """
    pass

def scroll_up() -> None:
    """Moves the window up by 100 lines.

    Args:
        None
    """
    pass

def create_file(filename: str) -> None:
    """Creates and opens a new file with the given name.

    Args:
        filename: str: The name of the file to create.
    """
    pass

def edit_file(start: int, end: int, content: str) -> None:
    """Edit a file.

    It replaces lines `start` through `end` (inclusive) with the given text `content` in the open file. Remember, the file must be open before editing.

    Args:
        start: int: The start line number. Must satisfy start >= 1.
        end: int: The end line number. Must satisfy start <= end <= number of lines in the file.
        content: str: The content to replace the lines with.
    """
    pass

def search_dir(search_term: str, dir_path: str = './') -> None:
    """Searches for search_term in all files in dir. If dir is not provided, searches in the current directory.

    Args:
        search_term: str: The term to search for.
        dir_path: Optional[str]: The path to the directory to search.
    """
    pass

def search_file(search_term: str, file_path: Optional[str] = None) -> None:
    """Searches for search_term in file. If file is not provided, searches in the current open file.

    Args:
        search_term: str: The term to search for.
        file_path: Optional[str]: The path to the file to search.
    """
    pass

def find_file(file_name: str, dir_path: str = './') -> None:
    """Finds all files with the given name in the specified directory.

    Args:
        file_name: str: The name of the file to find.
        dir_path: Optional[str]: The path to the directory to search.
    """
    pass

def parse_pdf(file_path: str) -> None:
    """Parses the content of a PDF file and prints it.

    Args:
        file_path: str: The path to the file to open.
    """
    pass

def parse_docx(file_path: str) -> None:
    """
    Parses the content of a DOCX file and prints it.

    Args:
        file_path: str: The path to the file to open.
    """
    pass

def parse_latex(file_path: str) -> None:
    """
    Parses the content of a LaTex file and prints it.

    Args:
        file_path: str: The path to the file to open.
    """
    pass

def parse_audio(file_path: str, model: str = 'whisper-1') -> None:
    """
    Parses the content of an audio file and prints it.

    Args:
        file_path: str: The path to the audio file to transcribe.
        model: Optional[str]: The audio model to use for transcription. Defaults to 'whisper-1'.
    """
    pass


def parse_image(
    file_path: str, task: str = 'Describe this image as detail as possible.'
) -> None:
    """
    Parses the content of an image file and prints the description.

    Args:
        file_path: str: The path to the file to open.
        task: Optional[str]: The task description for the API call. Defaults to 'Describe this image as detail as possible.'.
    """
    pass


def parse_video(
    file_path: str,
    task: str = 'Describe this image as detail as possible.',
    frame_interval: int = 30,
) -> None:
    """
    Parses the content of an image file and prints the description.

    Args:
        file_path: str: The path to the video file to open.
        task: Optional[str]: The task description for the API call. Defaults to 'Describe this image as detail as possible.'.
        frame_interval: Optional[int]: The interval between frames to analyze. Defaults to 30.

    """
    pass

def parse_pptx(file_path: str) -> None:
    """
    Parses the content of a pptx file and prints it.

    Args:
        file_path: str: The path to the file to open.
    """
    pass
\end{minted}

\section{BrowserGym Actions}
\label{sec:browsergym-actions}
The following are all the supported actions defined in BrowserGym\footnote{\url{https://github.com/ServiceNow/BrowserGym/blob/main/core/src/browsergym/core/action/functions.py}} as of v0.3.4.
The actions can be categorized into several types and can be configured to use only a subset of the functionality.
There are agent control actions, navigation actions, page element-based actions, coordinate-based actions, as well as tab-related actions.
We use these actions from the BrowserGym library as our main browsing action primitives.

\begin{minted}[breaklines]{python}
def send_msg_to_user(text: str):
    """
    Sends a message to the user.

    Examples:
        send_msg_to_user("Based on the results of my search, the city was built in 1751.")
    """
    pass

def report_infeasible(reason: str):
    """
    Notifies the user that their instructions are infeasible.

    Examples:
        report_infeasible("I cannot follow these instructions because there is no email field in this form.")
    """
    pass


def noop(wait_ms: float = 1000):
    """
    Do nothing, and optionally wait for the given time (in milliseconds).

    Examples:
        noop()
        noop(500)
    """
    pass


# https://playwright.dev/docs/input#text-input
def fill(bid: str, value: str):
    """
    Fill out a form field. It focuses the element and triggers an input event with the entered text.
    It works for <input>, <textarea> and [contenteditable] elements.

    Examples:
        fill('237', 'example value')
        fill('45', "multi-line\\nexample")
        fill('a12', "example with \\"quotes\\"")
    """
    pass


# https://playwright.dev/python/docs/api/class-locator#locator-check
def check(bid: str):
    """
    Ensure a checkbox or radio element is checked.

    Examples:
        check('55')
    """
    pass


# https://playwright.dev/python/docs/api/class-locator#locator-uncheck
def uncheck(bid: str):
    """
    Ensure a checkbox or radio element is unchecked.

    Examples:
        uncheck('a5289')
    """
    pass


# https://playwright.dev/docs/input#select-options
def select_option(bid: str, options: str | list[str]):
    """
    Select one or multiple options in a <select> element. You can specify
    option value or label to select. Multiple options can be selected.

    Examples:
        select_option('a48', "blue")
        select_option('c48', ["red", "green", "blue"])
    """
    pass


# https://playwright.dev/python/docs/api/class-locator#locator-click
def click(
    bid: str,
    button: Literal["left", "middle", "right"] = "left",
    modifiers: list[Literal["Alt", "Control", "Meta", "Shift"]] = [],
):
    """
    Click an element.

    Examples:
        click('a51')
        click('b22', button="right")
        click('48', button="middle", modifiers=["Shift"])
    """
    pass


# https://playwright.dev/python/docs/api/class-locator#locator-dblclick
def dblclick(
    bid: str,
    button: Literal["left", "middle", "right"] = "left",
    modifiers: list[Literal["Alt", "Control", "Meta", "Shift"]] = [],
):
    """
    Double click an element.

    Examples:
        dblclick('12')
        dblclick('ca42', button="right")
        dblclick('178', button="middle", modifiers=["Shift"])
    """
    pass


# https://playwright.dev/python/docs/api/class-locator#locator-hover
def hover(bid: str):
    """
    Hover over an element.

    Examples:
        hover('b8')
    """
    pass


# https://playwright.dev/python/docs/input#keys-and-shortcuts
def press(bid: str, key_comb: str):
    """
    Focus the matching element and press a combination of keys. It accepts
    the logical key names that are emitted in the keyboardEvent.key property
    of the keyboard events: Backquote, Minus, Equal, Backslash, Backspace,
    Tab, Delete, Escape, ArrowDown, End, Enter, Home, Insert, PageDown, PageUp,
    ArrowRight, ArrowUp, F1 - F12, Digit0 - Digit9, KeyA - KeyZ, etc. You can
    alternatively specify a single character you'd like to produce such as "a"
    or "#". Following modification shortcuts are also supported: Shift, Control,
    Alt, Meta.

    Examples:
        press('88', 'Backspace')
        press('a26', 'Control+a')
        press('a61', 'Meta+Shift+t')
    """
    pass


# https://playwright.dev/python/docs/api/class-locator#locator-focus
def focus(bid: str):
    """
    Focus the matching element.

    Examples:
        focus('b455')
    """
    pass


# https://playwright.dev/python/docs/api/class-locator#locator-clear
def clear(bid: str):
    """
    Clear the input field.

    Examples:
        clear('996')
    """
    pass


# https://playwright.dev/python/docs/input#drag-and-drop
def drag_and_drop(from_bid: str, to_bid: str):
    """
    Perform a drag & drop. Hover the element that will be dragged. Press
    left mouse button. Move mouse to the element that will receive the
    drop. Release left mouse button.

    Examples:
        drag_and_drop('56', '498')
    """
    pass


# https://playwright.dev/python/docs/api/class-mouse#mouse-wheel
def scroll(delta_x: float, delta_y: float):
    """
    Scroll horizontally and vertically. Amounts in pixels, positive for right or down scrolling, negative for left or up scrolling. Dispatches a wheel event.

    Examples:
        scroll(0, 200)
        scroll(-50.2, -100.5)
    """
    pass


# https://playwright.dev/python/docs/api/class-mouse#mouse-move
def mouse_move(x: float, y: float):
    """
    Move the mouse to a location. Uses absolute client coordinates in pixels.
    Dispatches a mousemove event.

    Examples:
        mouse_move(65.2, 158.5)
    """
    pass


# https://playwright.dev/python/docs/api/class-mouse#mouse-up
def mouse_up(x: float, y: float, button: Literal["left", "middle", "right"] = "left"):
    """
    Move the mouse to a location then release a mouse button. Dispatches
    mousemove and mouseup events.

    Examples:
        mouse_up(250, 120)
        mouse_up(47, 252, 'right')
    """
    pass


# https://playwright.dev/python/docs/api/class-mouse#mouse-down
def mouse_down(x: float, y: float, button: Literal["left", "middle", "right"] = "left"):
    """
    Move the mouse to a location then press and hold a mouse button. Dispatches
    mousemove and mousedown events.

    Examples:
        mouse_down(140.2, 580.1)
        mouse_down(458, 254.5, 'middle')
    """
    pass


# https://playwright.dev/python/docs/api/class-mouse#mouse-click
def mouse_click(x: float, y: float, button: Literal["left", "middle", "right"] = "left"):
    """
    Move the mouse to a location and click a mouse button. Dispatches mousemove,
    mousedown and mouseup events.

    Examples:
        mouse_click(887.2, 68)
        mouse_click(56, 712.56, 'right')
    """
    pass


# https://playwright.dev/python/docs/api/class-mouse#mouse-dblclick
def mouse_dblclick(x: float, y: float, button: Literal["left", "middle", "right"] = "left"):
    """
    Move the mouse to a location and double click a mouse button. Dispatches
    mousemove, mousedown and mouseup events.

    Examples:
        mouse_dblclick(5, 236)
        mouse_dblclick(87.5, 354, 'right')
    """
    pass


def mouse_drag_and_drop(from_x: float, from_y: float, to_x: float, to_y: float):
    """
    Drag and drop from a location to a location. Uses absolute client
    coordinates in pixels. Dispatches mousemove, mousedown and mouseup
    events.

    Examples:
        mouse_drag_and_drop(10.7, 325, 235.6, 24.54)
    """
    pass


# https://playwright.dev/python/docs/api/class-keyboard#keyboard-press
def keyboard_press(key: str):
    """
    Press a combination of keys. Accepts the logical key names that are
    emitted in the keyboardEvent.key property of the keyboard events:
    Backquote, Minus, Equal, Backslash, Backspace, Tab, Delete, Escape,
    ArrowDown, End, Enter, Home, Insert, PageDown, PageUp, ArrowRight,
    ArrowUp, F1 - F12, Digit0 - Digit9, KeyA - KeyZ, etc. You can
    alternatively specify a single character you'd like to produce such
    as "a" or "#". Following modification shortcuts are also supported:
    Shift, Control, Alt, Meta.

    Examples:
        keyboard_press('Backspace')
        keyboard_press('Control+a')
        keyboard_press('Meta+Shift+t')
        page.keyboard.press("PageDown")
    """
    pass


# https://playwright.dev/python/docs/api/class-keyboard#keyboard-up
def keyboard_up(key: str):
    """
    Release a keyboard key. Dispatches a keyup event. Accepts the logical
    key names that are emitted in the keyboardEvent.key property of the
    keyboard events: Backquote, Minus, Equal, Backslash, Backspace, Tab,
    Delete, Escape, ArrowDown, End, Enter, Home, Insert, PageDown, PageUp,
    ArrowRight, ArrowUp, F1 - F12, Digit0 - Digit9, KeyA - KeyZ, etc.
    You can alternatively specify a single character you'd like to produce
    such as "a" or "#".

    Examples:
        keyboard_up('Shift')
        keyboard_up('c')
    """
    pass


# https://playwright.dev/python/docs/api/class-keyboard#keyboard-down
def keyboard_down(key: str):
    """
    Press and holds a keyboard key. Dispatches a keydown event. Accepts the
    logical key names that are emitted in the keyboardEvent.key property of
    the keyboard events: Backquote, Minus, Equal, Backslash, Backspace, Tab,
    Delete, Escape, ArrowDown, End, Enter, Home, Insert, PageDown, PageUp,
    ArrowRight, ArrowUp, F1 - F12, Digit0 - Digit9, KeyA - KeyZ, etc. You can
    alternatively specify a single character such as "a" or "#".

    Examples:
        keyboard_up('Shift')
        keyboard_up('c')
    """
    pass


# https://playwright.dev/python/docs/api/class-keyboard#keyboard-type
def keyboard_type(text: str):
    """
    Types a string of text through the keyboard. Sends a keydown, keypress/input,
    and keyup event for each character in the text. Modifier keys DO NOT affect
    keyboard_type. Holding down Shift will not type the text in upper case.

    Examples:
        keyboard_type('Hello world!')
    """
    pass


# https://playwright.dev/python/docs/api/class-keyboard#keyboard-insert-text
def keyboard_insert_text(text: str):
    """
    Insert a string of text in the currently focused element. Dispatches only input
    event, does not emit the keydown, keyup or keypress events. Modifier keys DO NOT
    affect keyboard_insert_text. Holding down Shift will not type the text in upper
    case.

    Examples:
        keyboard_insert_text('Hello world!')
    """
    pass


# https://playwright.dev/python/docs/api/class-page#page-goto
def goto(url: str):
    """
    Navigate to a url.

    Examples:
        goto('http://www.example.com')
    """
    pass


# https://playwright.dev/python/docs/api/class-page#page-go-back
def go_back():
    """
    Navigate to the previous page in history.

    Examples:
        go_back()
    """
    pass


# https://playwright.dev/python/docs/api/class-page#page-go-forward
def go_forward():
    """
    Navigate to the next page in history.

    Examples:
        go_forward()
    """
    pass


# https://playwright.dev/python/docs/api/class-browsercontext#browser-context-new-page
def new_tab():
    """
    Open a new tab. It will become the active one.

    Examples:
        new_tab()
    """
    global page
    # set the new page as the active page
    page = page.context.new_page()
    # trigger the callback that sets this page as active in browsergym
    pass


# https://playwright.dev/python/docs/api/class-page#page-close
def tab_close():
    """
    Close the current tab.

    Examples:
        tab_close()
    """
    pass


# https://playwright.dev/python/docs/api/class-page#page-bring-to-front
def tab_focus(index: int):
    """
    Bring tab to front (activate tab).

    Examples:
        tab_focus(2)
    """
    pass


# https://playwright.dev/python/docs/input#upload-files
def upload_file(bid: str, file: str | list[str]):
    """
    Click an element and wait for a "filechooser" event, then select one
    or multiple input files for upload. Relative file paths are resolved
    relative to the current working directory. An empty list clears the
    selected files.

    Examples:
        upload_file("572", "my_receipt.pdf")
        upload_file("63", ["/home/bob/Documents/image.jpg", "/home/bob/Documents/file.zip"])
    """
    pass


# https://playwright.dev/python/docs/input#upload-files
def mouse_upload_file(x: float, y: float, file: str | list[str]):
    """
    Click a location and wait for a "filechooser" event, then select one
    or multiple input files for upload. Relative file paths are resolved
    relative to the current working directory. An empty list clears the
    selected files.

    Examples:
        mouse_upload_file(132.1, 547, "my_receipt.pdf")
        mouse_upload_file(328, 812, ["/home/bob/Documents/image.jpg", "/home/bob/Documents/file.zip"])
    """
    pass

\end{minted}

\section{Browsing Agent Details}
\label{sec:browsing-agent-details}

The following shows an example prompt containing all the information required for the current step to make a prediction about the next browsing actions.
Note that we also instruct the agent to predict multiple actions in one turn if the agent thinks they are meant to be executed sequentially without any feedback from the page.
This could save turns for common workflows that consist of a sequence of actions on the same page without any observation change, such as filling the username and password and submit in a login page.

\begin{minted}[breaklines]{text}
# Instructions
Review the current state of the page and all other information to find the best possible next action to accomplish your goal. Your answer will be interpreted and executed by a program, make sure to follow the formatting instructions.

# Goal:
Browse localhost:8000, and tell me the ultimate answer to life. Do not ask me for confirmation at any point.

# Action Space

16 different types of actions are available.

noop(wait_ms: float = 1000)
    Examples:
        noop()

        noop(500)

send_msg_to_user(text: str)
    Examples:
        send_msg_to_user('Based on the results of my search, the city was built in 1751.')

scroll(delta_x: float, delta_y: float)
    Examples:
        scroll(0, 200)

        scroll(-50.2, -100.5)

fill(bid: str, value: str)
    Examples:
        fill('237', 'example value')

        fill('45', 'multi-line\nexample')

        fill('a12', 'example with "quotes"')

select_option(bid: str, options: str | list[str])
    Examples:
        select_option('48', 'blue')

        select_option('48', ['red', 'green', 'blue'])

click(bid: str, button: Literal['left', 'middle', 'right'] = 'left', modifiers: list[typing.Literal['Alt', 'Control', 'Meta', 'Shift']] = [])
    Examples:
        click('51')

        click('b22', button='right')

        click('48', button='middle', modifiers=['Shift'])

dblclick(bid: str, button: Literal['left', 'middle', 'right'] = 'left', modifiers: list[typing.Literal['Alt', 'Control', 'Meta', 'Shift']] = [])
    Examples:
        dblclick('12')

        dblclick('ca42', button='right')

        dblclick('178', button='middle', modifiers=['Shift'])

hover(bid: str)
    Examples:
        hover('b8')

press(bid: str, key_comb: str)
    Examples:
        press('88', 'Backspace')

        press('a26', 'Control+a')

        press('a61', 'Meta+Shift+t')

focus(bid: str)
    Examples:
        focus('b455')

clear(bid: str)
    Examples:
        clear('996')

drag_and_drop(from_bid: str, to_bid: str)
    Examples:
        drag_and_drop('56', '498')

upload_file(bid: str, file: str | list[str])
    Examples:
        upload_file('572', 'my_receipt.pdf')

        upload_file('63', ['/home/bob/Documents/image.jpg', '/home/bob/Documents/file.zip'])

go_back()
    Examples:
        go_back()

go_forward()
    Examples:
        go_forward()

goto(url: str)
    Examples:
        goto('http://www.example.com')

Multiple actions can be provided at once. Example:
fill('a12', 'example with "quotes"')
click('51')
click('48', button='middle', modifiers=['Shift'])
Multiple actions are meant to be executed sequentially without any feedback from the page.
Don't execute multiple actions at once if you need feedback from the page.

# Current Accessibility Tree:
RootWebArea 'The Ultimate Answer', focused
	[8] heading 'The Ultimate Answer'
	[9] paragraph ''
		StaticText 'Click the button to reveal the answer to life, the universe, and everything.'
	[10] button 'Click me', clickable

# Previous Actions
goto('http://localhost:8000')

Here is an example with chain of thought of a valid action when clicking on a button:
"
In order to accomplish my goal I need to click on the button with bid 12
```click("12")```
\end{minted}

And an example response to the above prompt is:

\begin{minted}[breaklines]{text}
In order to accomplish my goal, I need to click on the button with bid 10 to reveal the answer to life, the universe, and everything.
```click("10")```
\end{minted}

For the evaluation on WebArena benchmark, since some of the tasks require checking for answer exact match on the agent's message back to the user, we add the following instruction to let the agent reply with only a concise answer string when messaging the user to prevent the agent from failing the test due to extra text:

\begin{minted}[breaklines]{text}
Here is another example with chain of thought of a valid action when providing a concise answer to user:
"
In order to accomplish my goal I need to send the information asked back to the user. This page list the information of HP Inkjet Fax Machine, which is the product identified in the objective. Its price is $279.49. I will send a message back to user with the answer.
```send_msg_to_user("$279.49")```
"
\end{minted}

\end{document}